\documentclass[preprint2]{aastex}
\usepackage{times}
\usepackage{amsmath}

\usepackage{chngcntr}

\newcommand{\mj}{$M_{\mathrm{J}}$}

\newcommand{\me}{$M_{\oplus}$}

\newcommand{\cp}{\citep}
\newcommand{\ct}{\citet}

\newcommand{\teff}{$T_{\rm eff}$}

\newcommand{\cas}{\emph{Cassini}}

\newenvironment{packed_enum}{
\begin{itemize}
  \setlength{\itemsep}{1pt}
  \setlength{\parskip}{0pt}
  \setlength{\parsep}{0pt}
}{\end{itemize}}

\begin{document}

\counterwithin{figure}{section}
\counterwithin{table}{section}
\numberwithin{equation}{section}

\setcounter{section}{3}

\title{\textbf{\LARGE 3}}
\title{\textbf{\LARGE  }}
\title{\textbf{\LARGE The Interior of Saturn}}

\author {\large Jonathan J. Fortney, Ravit Helled, Nadine Nettelmann, \\ David J. Stevenson, Mark S. Marley, William B. Hubbard, Luciano Iess}

\begin{abstract}
\baselineskip = 11pt
\leftskip = 0.65in 
\rightskip = 0.65in 
\parindent=1pc
\end{abstract}  

{\emph{\normalsize Copyright Notice}}\\
\emph{``The Chapter, 'The Interior of Saturn', is to be published by
Cambridge University Press as part of a multivolume
work edited by Kevin Baines, Michael Flasar,
Norbert Krupp, and Thomas Stallard, entitled
``Saturn in the 21st Century'' ('the Volume')
\\
\textcopyright in the Chapter, Jonathan J. Fortney, Ravit Helled, Nadine Nettelmann, David J. Stevenson, Mark S. Marley, William B. Hubbard, Luciano Iess \textcopyright in the Volume,
Cambridge University Press
\\
NB: The copy of the Chapter, as displayed on this
website, is a draft, pre-publication copy only. The
final, published version of the Chapter will be
available to purchase through Cambridge University
Press and other standard distribution channels as
part of the wider, edited Volume, once published.
This draft copy is made available for personal use
only and must not be sold or re-distributed.''}
\\
\\
{\textbf{\normalsize Abstract}} \\We review our current understanding of the interior structure and thermal evolution of Saturn, with a focus on recent results in the \cas\ era.  There has been important progress in understanding physical inputs, including equations of state of planetary materials and their mixtures, physical parameters like the gravity field and rotation rate, and constraints on Saturnian free oscillations.  At the same time, new methods of calculation, including work on the gravity field of rotating fluid bodies, and the role of interior composition gradients, should help to better constrain the state of Saturn's interior, now and earlier in its history.  However, a better appreciation of modeling uncertainties and degeneracies, along with a greater exploration of modeling phase space, still leave great uncertainties in our understanding of Saturn's interior.  Further analysis of \cas\ data sets, as well as precise gravity field measurements from the \cas\ Grand Finale orbits, will further revolutionize our understanding of Saturn's interior over the next few years.

\subsection{Introduction}  
In investigations into the interior structure, composition, and thermal evolution of giant planets, Saturn can sometimes receive ``Second City'' status compared to the bright lights of Jupiter.  Both planets are natural laboratories for understanding the physics of hydrogen, helium, and their mixtures, under high pressure.  Since both planets are predominantly composed of H/He, understanding their compositions sheds important and unique light on the composition of the solar nebula during the era of planet formation.

While Jupiter is often thought of as the benchmark giant planet for this class of astrophysical object, now known to be abundant in the universe, Saturn provides an interesting point of comparison and departure for understanding giant planet structure and evolution.  For instance, Jupiter models are highly sensitive to the equation of state (EOS, the relation between important quantities such as temperature, pressure, and density) of hydrogen, the most abundant element in the universe, and thus can help to probe the phase space region around a few megabars and ten thousand Kelvin, for which accurate lab experimental data are not available yet.  Saturn on the other hand, with 30\% of Jupiter's mass, probes less of hydrogen's phase space, but has its own host of complex issues.  With its peculiar magnetic field and high intrinsic luminosity, Saturn provides challenges to our understanding of the first-order properties that define a gas giant planet.  For both planets, an understanding of their bulk composition can come from interior modeling, which is an important constraint on formation scenarios.

In looking back at the post-\emph{Voyager} Saturn review chapter of \ct{Hubbard84}, it is apparent that a number of the important issues of the day are still unsolved.  What is the enrichment of heavy elements compared to the Sun, and Jupiter, and what is their distribution within the planet?  What is the mass of any heavy element core?  To what degree has the phase separation of helium in the planet's deep interior altered the evolutionary history of the planet?  Are there deviations from adiabaticity? 

Understanding the interior of Saturn crosses diverse fields from condensed matter physics to planet formation, but progress is challenging due to uncertainties in input physics as well as in the indirect nature of our constraints on the planet.  This era near the end of the \cas\ Mission is an excellent time to review our understanding of Saturn's interior.  We are at a time where new observational constraints, such as a refined measurement of the gravity field as well as ring seismology, new theoretical and experimental constraints on input physics like the hydrogen-helium phase diagram, and new methods of calculating interior models, 
are all coming together to allow for a new understanding of Saturn's interior, and by extension the interiors of giant planets as a class of astrophysical objects.

\subsection{Available data and its applications}

\subsubsection{Energy balance} 
Like Jupiter, the power incident upon Saturn due to solar radiation is on the same order as the intrinsic power from the planet.  The thermal flux detected from the planet today is a combination of this intrinsic flux, which is a remnant of formation, as well as thermalized solar energy.  To distinguish between these components, the Bond albedo of Saturn must be determined, as this quantity yields the total flux absorbed by the planet, which is then re-radiated.  Models aim to understand the intrinsic flux from the planet's interior, and how the planet cooled to this flux level at 4.56 Gyr \cp[e.g.][]{Fortney11}.  In addition the 1-bar temperature dictates the specific entropy of an isentropic deep interior (see Section \ref{assumption}), setting the upper boundary for the thermal structure of the interior.  A self-consistent model should fit the intrinsic flux as well as the 1-bar temperature, in addition to other constraints detailed below.  Table \ref{tab:balance} shows the current energy budget of the planet.

 \begin{table*}
 \centering
 \begin{tabular}{||c c c c c c c||}
 \hline
Absorbed & Emitted & Intrinsic & Intrinsic & Bond & \teff & $T_{\rm 1 bar}$\\  
Power & Power & Power & Flux & Albedo &  &  \\
$10^{23}$ erg s$^{-1}$ & $10^{23}$ erg s$^{-1}$ & $10^{23}$ erg s$^{-1}$ & erg s$^{-1}$ cm$^{-2}$ &  & K & K \\
 \hline
 11.14(50) & 19.77(32) & 8.63(60) & 2010(140) & 0.342(30) & 95.0(4) & 135(5) \\ 
 \hline
 \end{tabular}
 \caption{Classic Values for Saturn Heat Balance:  All data after \ct{Pearl91} except $T_{\rm 1 bar}$, from \ct{Lindal85, Lindal92}. It should be noted that the radio occultation retrieved profile depends on the atmospheric composition which was assumed to be 94\% (by number) of molecular hydrogen with the rest being helium.  Furthermore, analysis of \emph{Cassini} data by \ct{Li10} yields revised, appreciably higher, values for the intrinsic flux and for the \teff, found to be $96.67 \pm 0.17$ K.}
  \label{tab:balance}
\end{table*}

\subsubsection{Atmospheric composition} 
Saturn's atmospheric composition is an important constraint on its interior structure and formation history (see the chapter by Atreya et al). In particular, if the H-He envelope
is fully convective and well-mixed, atmospheric abundances that can be measured either spectroscopically or \emph{in situ} should be representative of the entire H-He envelope.  This could yield complementary information on the heavy element enrichment of the H-He evelope that would be distinct from that of the gravity field.  While Jupiter's atmosphere above 22 bars was directly sampled by the \emph{Galileo Entry Probe}, no similar probe for Saturn is currently scheduled.  Jupiter's atmosphere is enhanced (by number) by factors of $\sim$2-5 relative to abundances in the Sun in most elements, with depletions in helium and neon attributed to interior processes (discussed below).  The well-known depletion in water is still an area of active discussion concerning whether the depletion reflects the true abundance of Jovian water \cp{Showman98,Lodders04,Mousis12,Helled14b}, which will hopefully be settled by the \emph{Juno} mission.

In terms of comparative planetary science, the only elemental abundance that has been accurately determined for each of the solar system's four giant planets is that of carbon, found in methane, where the super-solar enhancement grows with decreasing planet mass, from a factor of $\sim$4 for Jupiter \cp{Wong04}, $\sim10$ for Saturn \cp{Fletcher09}, and $\sim$80 for Uranus and Neptune \cp{Sromovsky11,Karkoschka11}.  For Saturn, this suggests a H-He envelope that on the whole may be strongly enhanced in heavy elements.  It has been problematic to directly include an implementation of an enriched envelope as a constraint in interior modeling, however.  One would also like to know the water and ammonia abundances, as these, along with methane, would likely correspond to $\sim$60-80 \% of the heavy element mass.  Given the \emph{Galileo} results, it may be very unwise to assume that O and N scale with C in giant planets.  If one knew the ``metallicity'' of the envelope of Saturn, that would place important constraints on the core mass, as a wide range of solutions for the bulk abundance of heavy elements allowed by the gravity field find a wide diversity of the amount of heavy elements in the envelope.  The abundance of helium is an essential constraint on interior models, as it affects the planet's density and temperature distribution with radius, as well as the planet's thermal evolution, and is discussed in detail later in the chapter.

\subsubsection{Gravity Field} 
The mass of Saturn is obtained from the observation of the motions of natural satellites: 95.161 \me, where 1 \me $= 5.97369 \times 10^{24}$ kg.  More precise measurements of the planet's gravity field can be obtained
through the analysis of the trajectories of spacecraft (e.g., \emph{Voyager}, \cas) during flyby (obtained via Doppler shift of radio emission).  The most precise constraints come from close-in passes to the planet, in a near-polar orbit.  Because of the rapid rotation of Saturn, its gravitational field departs from that of a point mass (a purely spherical field).

Of particular interest for using the gravity field is the need for a suitable theory to invert the gravity information to provide constraints on the planet's density as a function of radius.  This ``Theory of Figures'' is a classical problem \cp{ZT78}, which is discussed in Section \ref{cms}.  For Saturn a limitation in the application of this theory is our uncertainty in the rotation rate of the planet, which is discussed in detail in Section \ref{rotate}.

\subsubsection{Magnetic field} \label{sec:mag} 
All of the solar system planets and some moons have or had large scale magnetic fields during their history, with Venus as the only possible large exception.  Six out of eight planets in our Solar system have present-day planetary-scale magnetic fields of internal origin, and all of the giant planets have large fields \cp[e.g.,][]{Stevenson03}. The planetary magnetic fields are as diverse as the host planets, yet no simple correlations have been found between the basic features of the magnetic fields (e.g., field strength, field morphology) and the basic features of the host planets (e.g., composition, mass, radius, rotation, heat flux). It is striking that Jupiter and Saturn, planets of a similar kind, should have such different fields, and this remains one of the biggest challenges to our understanding.  (See the chapter by Christensen et al.)

In-situ magnetic field measurements made by \emph{Pioneer 11} Saturn flyby in 1979 showed for the first time the existence of a dipole-dominant, global-scale magnetic field at Saturn with surface field strength around 30,000 nT \cp{Smith80}.  Subsequent MAG measurements made during the \emph{Voyager 1}, \emph{Voyager 2} Saturn flybys \cp{Ness81,Ness82} and those made with the ongoing \emph{Cassini} orbital mission \cp{Dougherty05,Burton09,Sterenborg10,Cao11,Cao12} have established the low degree structures of Saturn's magnetic field.  \ct{Cao11,Cao12} employed spherical harmonic analysis based on the close-in part of the \cas\ MAG measurements from Saturn orbital insertion (SOI) to early 2010 and showed that Saturn's magnetic field is extremely axisymmetric with an upper bound on its dipole tilt of 0.06 degrees. This is in striking contrast to the Earthlike dipole tilt of ten degrees exhibited by Jupiter. It also explains the persistent uncertainty in the spin rate of Saturn (discussed in section \ref{rotate}), whose value for a fluid planet can probably only be meaningfully defined by consideration of the magnetic field non-spinaxisymmetry.

One cannot rule out for certain the presence of a non-axisymmetric gravity field component, the value of which might need only be one part in $\sim$~$10^8$ or even smaller to be detectable. 
However, there is no assurance that this would represent rotation of the deep interior, whereas the magnetic field deep down is prevented from having a significant differential rotation because of the large toroidal field and resulting torques that would otherwise result. Saturn also has a modest north-south asymmetry with an axial quadrupole moment that amounts to 7.5\% of the axial dipole moment on the surface, and has extremely slow time evolution between the \emph{Cassini}-era and the \emph{Pioneer-Voyager}-era, with an upper bound one order of magnitude smaller than that of the geomagnetic field.

The external magnetic field $\boldsymbol B$ is often expressed as an expansion in spherical harmonics of the scalar potential $W$, with ${\boldsymbol B}= - \nabla W$:
\begin{equation}
\begin{split}
W = R_{\mathrm eq} \sum_{n=1}^\infty \left( \frac{R_{\rm eq}}{r} \right)^{n+1} \sum_{m=0}^{n} \\ \left(g_n^m \cos{(m \phi)}  + 
h_n^m \sin{(m \phi)}\right) P_n^m \cos{(\theta)},
\label{W}
\end{split}
\end{equation}

where $\phi$ is the longitude and the $P_n^m$ terms are the associated Legendre polynomials.  The coefficients $g_n^m$ and $h_n^m$ are the magnetic moments that characterize the field, and are in units of Tesla.

Since the field is indistinguishable from axisymmetry in the current data, it suffices to list the axial terms in the usual expansion of the potential whose gradient provides the field. Given here in Table 2 is the \emph{Cassini} 5 model of \ct{Cao12}. As is usual in field models, one cannot exclude higher harmonics at level of tens of nT and accordingly the uncertainties in the listed values are of that order.

\begin{table}
\centering
 \begin{tabular}{|c c|}
 \hline
Spherical Harmonic Coefficient & Amplitude (nT) \\
 \hline
$g_1^0$	&	21191 \\
$g_2^0$	&	1586 \\
$g_3^0$ & 	2374 \\
$g_4^0$	&	65 \\
$g_5^0$	&	185 \\
 \hline
 \end{tabular}
 \caption{Axial terms of \emph{Cassini} 5 magnetic field model of Cao et al. (2012).}
  \label{tab:mag}
\end{table}

As with Jupiter and Earth, the quadrupole is suppressed (that is, the octupole is larger) but since the field is so nearly axisymmetric, it is also evident that the field has very little radial flux near the equator (recall that the dipole has none and the octupole has none). This may be of significance for interpreting the behavior of the dynamo that produces the field.

It is common (though perhaps dangerous) practice to infer the size of the dynamo region by asking for the radius at which the field at higher harmonics approaches the dipole in magnitude. By this measure (and using the octupole as a better guide than the suppressed quadrupole) we can infer a ``core radius'' of $(2374 \times 4/21191 \times 2)^{1/2} \sim 0.47$ in units of Saturn's radius, not greatly different from the radius at which hydrogen becomes highly conducting within Saturn.  However, this should not be over-interpreted.

\subsection{Input physics} 
\subsubsection{Hydrogen}
Within giant planets, hydrogen is found in the fluid, not solid phase.  Most of the mass of Saturn is beyond the realm of current experiment on hydrogen, so a mix of constraints from lower pressure experiments, together with simulations of hydrogen under high pressure, are used to understand its EOS.  The EOS of hydrogen is the most important physical input into Saturn interior models.  However, uncertainties in the EOS for Saturn models are not as important as for Jupiter \cp{Saumon04} since Saturn is lower in mass, and therefore less of its interior is found in the higher pressure regions above several Mbar that is not yet accessible to experiment.

For a given interior isentrope, the Gr\"uneisen parameter indicates the change of temperature with density within the interior, and hence the bulk energy reservoir of a given model.  The compressibility of hydrogen, as a function of pressure (and hence, radius) directly affects inferences for the amount of heavy elements within the planet needed to explain its radius and gravity field, as well as \emph{where} these heavy elements are found within the planet -- meaning, perhaps within a core, which may not be fully distinct from the overlying envelope, or distributed within the H/He-dominated envelope.

The past 15 years has seen dramatic advances in our understanding of hydrogen, both in the realm of experiment and simulation.  On the experimental side, reverberation shock measurements of the conductivity of hydrogen at the fluid insulator (H$_2$) to fluid metal (H$^+$) transition indicate a continuous transition from the molecular to the metallic phase \cp{Nellis99}.  First-principles simulations of this transition also find a continuous transition over the temperatures of interest for giant planets \cp[see][for a comprehensive review]{McMahon12}.

\subsubsection{H/He mixtures}
How the physics of mixtures of hydrogen and helium may differ from a simple linear mixture of the two components has been an area of active study for decades.  A number of earlier investigations suggested the helium may phase separate from liquid metallic hydrogen under giant planet conditions and ``rain down'' to deeper layers within the planet \cp{Salpeter73,Stevenson75,SS77a,SS77b}.  Phase separation occurs when the Gibbs free energy of a mixture can be minimized when the mixture separates into two distinct phases -- here, where one is helium poor, and the other, helium rich.

\begin{figure}[h!] 
\centering
\includegraphics[width = 3.3in]{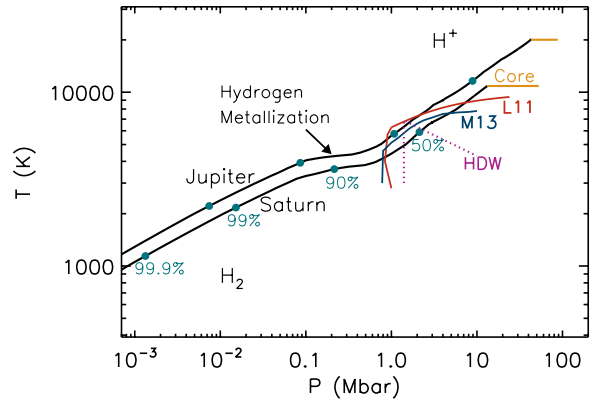}
\caption{Interior $P-T$ profiles of Jupiter and Saturn, following the methods of \ct{Nettelmann13b}.  The H/He envelopes are colored in black, with the (assumed) isothermal cores are in orange.  For both planets, blue-green dots indicate where the enclosed mass of the model planet is 50\%, 90\%, 99\%, and 99.9\%.  Note the shift outward to lower pressures at at given mass shell for Saturn, compared to Jupiter.  The gradual transition from fluid H$_2$ to liquid metallic hydrogen (H$^+$) is shown with a black arrow.  In red, blue, and purple are three predicted regions of He immiscibility (at $Y=0.27$, the protosolar abundance) in hydrogen.  The theory of \ct{HDW}, analogous to \ct{Stevenson75}, is labeled ``HDW.''  The theories of \ct{Lorenzen11} and \ct{Morales13} are labelled ``L11'' and ``M13'' respectively.  Both of these recent \emph{ab initio} simulation predict that large regions of Saturn's interior mass is within the He immiscibility region.}
\label{fig:profiles}
\end{figure}

Early work on trying to understand the phase diagram of H-He mixtures focused on systems that were readily amenable to calculation, for instance mixtures of fully ionized H and He \cp{Stevenson75,HDW}.  These calculations suggested that Saturn's current isentrope, and perhaps Jupiter's, intercepted regions of $P-T$ where phase separation would occur.  As \emph{ab initio} methods became possible, the phase diagram was investigated with these tools \cp{Klepeis91,Pfaff95}, but the results of these early 90's works were significantly inconsistent with each other.  With the rise of more modern \emph{ab initio} tools that were able to make fewer approximations, H-He mixtures have again been investigated.  Deviations from linear mixing have been found, and new calculations of the phase diagram have been published by two groups, \ct{Lorenzen09,Lorenzen11} and \ct{Morales09,Morales13}.  Isentropic interior models of Jupiter and Saturn, compared to these various H/He phase diagrams, are shown in Figure \ref{fig:profiles}.  All recent work finds that Saturn's estimated interior $P-T$ profile intersects regions of He phase separation.  He-rich droplets, being denser than their surroundings, may rain down to deeper layers of the planet, redistributing significant mass and altering the cooling history of the planet \cp{SS77a,FH03}.

Beside the onset temperature for immiscibility of the mixture, the shape of the phase diagram is important, as that directly controls the fraction of the planet's mass that falls within the phase separation region.  Earlier work \cp{Stevenson75,HDW} suggested He immiscibility in a relatively narrow pressure range, which grew slowly as the planet's interior cooled.  However, \ct{FH03} suggested that Saturn's current luminosity can be explained by settling of helium droplets throughout most of the planetary interior.  Modern phase diagrams \cp{Lorenzen11,Morales13} basically agree that this immiscibility region includes the bulk of Saturn's interior.

\subsubsection{Water and rock}
The EOS for the heavier elements have generally received somewhat less attention than those for hydrogen and helium.  However, the past five years has seen substantial advances in the \emph{ab initio} calculations of the EOS, as well as miscibility properties, for water, ammonia, rock, and iron.

Perhaps most importantly, an accurate \emph{ab initio} EOS for water has been published by \ct{French09}, which fares extremely well against data from single and double-shock experiments up to an impressive 7 Mbar \cp{Knudson12}.  The phase diagram of water has been explored, and it appears rather conclusive that any water in Saturn's core is found in the fluid, not solid state.  \ct{Wilson12a} have also looked at whether water, at Saturn and Jupiter's core conditions, is miscible in liquid metallic hydrogen.  They find that it is, such that for both planets diffusion of core material into the overlying H/He envelope is probable, although the efficiency of this process is still unknown. 

The details of the EOS of rock and iron are less essential, as the temperature dependence on the density of these component is relatively weak at giant planet interior conditions.  Approximate EOS for rock-iron mixtures can be found in \ct{Hubbard89}, and \ct{Saumon04} use a ``dry sand'' EOS from the Sesame database.  \ct{Wilson12b} and \ct{Wahl13} have looked at miscibility of silicate rock MgO, and iron, respectively and, like for water, find that these components are also miscible in liquid metallic hydrogen.  However, MgO was found to be solid under Saturnian conditions, while iron is likely solid today, but perhaps liquid at earlier times when the core was hotter.  
The miscibility behavior of core material is an important but by itself incomplete indicator for the efficiency of core erosion. Thus at present, the amount of possibly redistributed core material in Jupiter and Saturn is not known.

\subsection{First order deductions from simple models} 
\subsubsection{Heavy-element enrichment from M-R} 
\begin{figure}[h!]
\centering
\includegraphics[width = 3.5in]{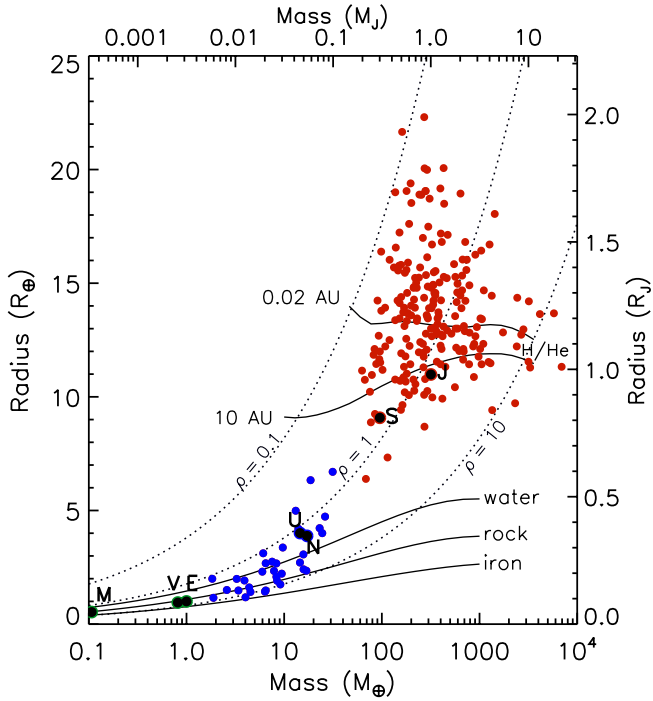}
\caption{The mass-radius relation of planets with well-determined masses, radii, and orbits.  Curves of constant bulk density ($\rho=0.1,1,10$) are shown as dotted lines.  Models for solar-composition planets at 4.5 Gyr, at 10 and 0.02 AU from the Sun, are shown as thick black curves \cp{Fortney07a}.  Jupiter, Saturn, Uranus, Neptune, Earth, and Venus are labeled by their first letter.  Planets more massive than 0.1 \mj\ are shown in red.  Saturn is appreciably lower that the 10 AU curve, indicating it is enriched in heavy elements.}
\label{fig:m-r}
\end{figure}

While there are many open questions regarding Saturn's bulk composition and internal structure, first-order knowledge of its composition can be inferred from the relation between its mass and radius (the Mass-Radius relation).  The Mass-Radius relation of planetary (or astrophysical) objects is often used to infer their bulk composition simply by 
%
calculating the mean density from $\bar {\rho}=3M/4\pi{\bar R}^3$ where $M$ is the planet's mass and $\bar{R}$ is its mean radius. 
As the mass of the object increases, the increase of density due to high pressures becomes important. 
Eventually, the density becomes so high with increasing planet mass that the radius will start to decrease. Such behavior is clearly seen in M-R relations for exoplanets, which exhibit a wide range of masses as shown in Fig.~\ref{fig:m-r}. However, inferring the bulk heavy element abundance of giant exoplanet from their M-R relation becomes challenging when their natural shrinking in radius over time is retarded by strong stellar irradiation. This can be the case for close-in exoplanets, e.g. on orbital distances of 0.02 AU. On the other hand, the radius 
of evolved planets at large orbital distances should be mainly affected by the bulk composition. 

From Figure \ref{fig:m-r} we  
can conclude that Saturn is enriched with heavy elements compared to proto-solar composition as its
radius is smaller than the one expected for a solar-composition planet at 10 AU. 
\par

\subsubsection{Dimensionless parameters}
There are several dimensionless parameters that are associated with the characterization of giant planets.  In particular, the rotation parameter $m$ or $q$, the nondimensional moment of inertia NMOI, and the flattening $f$ are linked by the Radau-Darwin approximation \citep{Jeffreys1924} and will be described in the following.
\\ 

\noindent {\bf Flattening (Oblateness):}

The flattening of a planet is defined by,
\begin{equation}
f\equiv\frac{R_{\rm eq}-R_{\rm p}}{R_{\rm eq}},
\end{equation}

where $R_{\rm eq}$ and $R_{\rm p}$ are the equatorial and polar radius, respectively. While knowledge of the continuous shape of a planet (i.e., radius vs.~latitude) is also desirable and available, typically, interior models use only the oblateness, or mean radius ${\bar R}$ that can be estimated from $\bar{R}\sim(R_{\rm eq}^2R_{\rm p})^{1/3}$. The oblateness provides information on the planetary rotation rate: planets that rotate rapidly are more oblate, as we will see below by using the Darwin-Radau relation.
\\

\noindent{\bf The rotation parameter:}

Typically, the density profile of giant planets is derived by using the theory of figures that is expanded in powers of a small parameter, the rotation parameter (e.g., Zharkov \& Trubitsyn, 1978), and discussed in Section \ref{cms}.
It is defined as the ratio of the centrifugal to gravitational force at the equator, $q\equiv \omega^2R_{\rm eq}^3/GM$, or alternatively with respect to the mean radius, $m\equiv \omega^2\bar{R}^3/GM$. The values for Saturn are $q\sim 0.155$ and $m\sim 0.139$ when using the \emph{Voyager} radio period. Since the rotation parameter depends on $\omega$ (i.e., the rotation rate) which is unknown for Saturn to within several minutes (see section \ref{rotate}) there is an uncertainty associated with the rotation parameter. For a rotation period that is ten minutes shorter than the \emph{Voyager} period the rotation parameters for Saturn are $q\sim 0.160$ and $m\sim 0.144$. The smaller the values of $q$ and $m$ are, the better is the approximation of the theory of figures (Hubbard, 2012).\\ 

\noindent {\bf Moment of inertia:}\\
The axial moment of inertia of a planetary body provides information on its density profile. 
Since giant planets are in hydrostatic equilibrium and therefore symmetric around the axis of rotation, their moment of inertia is derived from:
\begin{equation}
I= 2\pi\int_0^{\pi}\int_0^R d\vartheta\,dr\: \rho(\vartheta,r)r^4\sin^3\vartheta.	
\end{equation}
It is common to define the nondimensional moment of inertia factor (hereafter, NMOI) as $I/MR^2$.
Then, the NMOI can be directly linked to the density (radial) distribution.  An NMOI of a constant-density object is 0.4, lower NMOI values correspond to objects that are more centrally condensed, i.e., increase of density toward the center. Therefore, just like $J_{2n}$, which is defined below in \ref{cms}, the MOI can be used  as an independent constraint on the internal density distribution.
\\

\noindent {\bf The Radau-Darwin approximation:}\\
Finally, one can link the dimensionless parameters by using the Radau-Darwin approximation \cp[e.g.,][]{ZT78}. There are several forms for this approximation. 
One of them relates the planetary NMOI and $\Lambda_2\equiv J_2/q$. The Radau-Darwin formula suggests that there is a one-to-one correspondence between MOI and $\Lambda_2$ by,
\begin{equation}
{\rm MOI}= \frac{I}{MR^2}= \frac{2}{3}\left[1 - \frac{2}{5}\left( \frac{5}{(\Lambda_2+1)-1} \right)^{1/2} \right].
\end{equation} 
Another form of the Radau-Darwin relates the MOI with the flattening $f$ and rotation parameter $q$ via,   
\begin{equation}\label{c}
\frac I{MR^2}=\frac 23\left[ 1-\frac 25\left( \frac{5q}{2f}-1\right)
^{1/2}\right]  
\end{equation}
For Saturn, the Radau-Darwin relation suggests an NMOI of 0.220 \cp[e.g.,][]{Guillot14,Helled11a}. 
For comparison, Jupiter's NMOI is estimated to be $\sim$0.265 \citep{Jeffreys1924,Helled11b} therefore indicating that Saturn is more centrally condensed than Jupiter, potentially indicating that Saturn has a larger core mass. 

The Radau-Darwin approximation is quite good.  \ct{Helled11a} investigated the possible range of Saturn's MOI values accounting for the uncertainty in rotation period and internal structure. It was found that the MOI value can differ by up to 10\% from the value derived by the Radau-Darwin relation. A similar analysis was also done for Jupiter \cp{Helled11b} for which the MOI range was found to be 0.264 with an uncertainty of up to 6\%. 
\par

\subsection{Current Modeling Methods and Assumptions}
\subsubsection{Theory of the Gravity Field} \label{cms} 
In the theory of figures (TOF), one uses potential theory to solve for the structure of rotationally-distorted Saturn, assuming hydrostatic equilibrium for a given pressure-density
relation $P(\rho)$.  
The external gravitational potential of a rotating
planet in hydrostatic equilibrium is given by, 
\begin{equation}
\small{
V = \frac{G M}{r} \left( 1 - \sum_{n=1}^\infty \left( \frac{R_{\rm eq}}{r} \right)^{2 n} J_{2 n} P_{2 n} \left( \cos \theta  \right)  \right)\\
+\frac{1}{2} \omega^2 r^2 \sin^2 \theta,
\label{V}
}
\end{equation}
where (r, $\theta, \phi$) are spherical polar coordinates and $R_{\rm eq}$ is the equatorial radius. 
It can be shown \citep{ZT78}
that the potential can be expressed as a
double power-series expansion
in the dimensionless small parameter $m$ defined above. 
Each term $J_{2n}$ in the multipole
expansion for Saturn's external gravitational potential $V$
then has a coefficient whose further expansion
takes the form
\begin{equation} \label{J2nsum}
J_{2n} = m^n \sum_{t=0}^{\infty} \Lambda_{2n}^{(t)} m^t,
\end{equation}
where the dimensionless response coefficients $\Lambda_{2n}^{(t)}$ must be
obtained from the solution of a hierarchy of integrodifferential
equations.  Since Saturn's $m \sim 0.14$, the expansion does not converge rapidly.
In principle, for comparison with expected high-precision \cas\ measurements of Saturn's
$J_{2n}$, using the expansion method
one would need to derive all of the response coefficients for a test $P(\rho)$
out to terms $\sim m^9$!
This situation thus indicates the need for a nonperturbative approach.

The multipole coefficients $J_{2n}$ are measurable by fitting
a multiparameter model to spacecraft Doppler residuals.  However, the corresponding model values
of $J_{2n}$ for a specified $P(\rho)$ are not obtained by expanding in powers
of $m$, but rather are calculated directly using an iterative self-consistent
solution to a prescribed precision, usually $\sim 10^{-12}$.  Two algorithms for
nonperturbative calculations are available.  One method (J.~Wisdom, 1996, unpublished, available at: http://web.mit.edu/wisdom/www/interior.pdf) assumes that the
interior density distribution is a continuous function of position and can be expanded on
a set of polynomials.  The other method (called the CMS or concentric Maclaurin spheroid method)
represents the interior density distribution by a nested set of spheroids, each of constant
density \citep{Hubbard12, Hubbard13}.  A CMS model can be made to approach a continuous-density model
by increasing the number of spheroids, at the price of lengthier computations.

It is not widely appreciated that traditional TOF methods
employ a formally nonconvergent expansion attributed to Laplace.  The suspect expansion
is in fact intimately related to the standard $J_{2n}$ expansion of the external gravity potential. Although criticisms of the expansion have been published
over the years, e.g. \citet{Kong13}, it can be shown \citep{Hubbard14} that both Jupiter and Saturn are in the domain of $m$ where Laplace's ``swindle'' works exactly, or at least to a precision $\sim 10^{-12}$, more than adequate for
quantitative comparison within the expected precision of \cas\ measurements.  An earlier
proof of the validity of the traditional expansion is given by \citet{Wavre30}.

\begin{figure}[h!] 
\centering
\includegraphics[width = 3.3in]{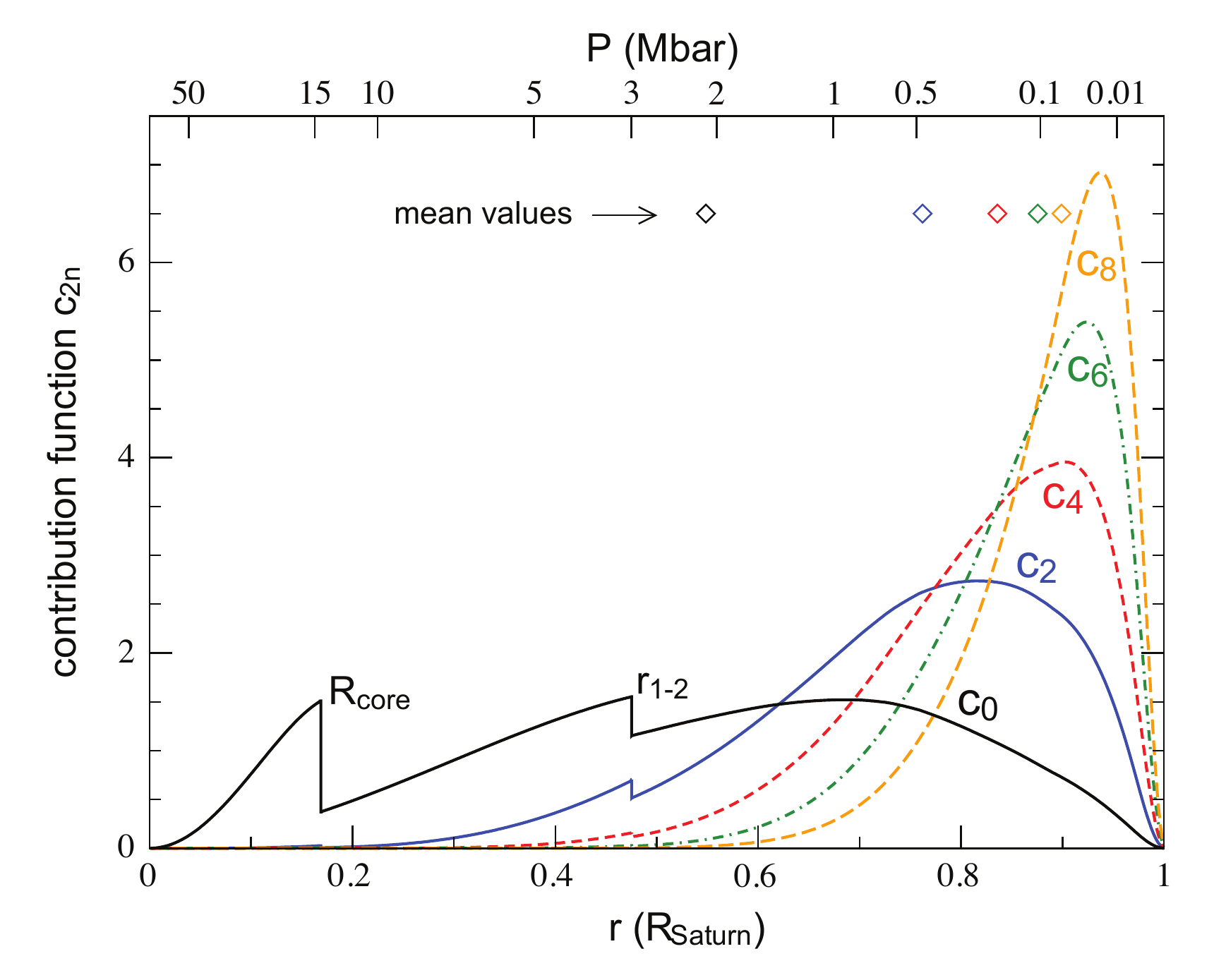}
\caption{Contribution functions of the gravitational harmonics $J_2$ (thick solid, blue), $J_4$ (short-dashed, red), $J_6$ (dot-dashed, green), and $J_8$ (long-dashed, orange) for the Saturn model ``S12-3a'' from \ct{Nettelmann13b}. The bottom x-axis is the fraction of Saturn's radius; the top x-axis shows the radii where the pressures of 0.01-50 Mbar occur. Layer boundaries due to the heavy element core, and between the helium-enriched interior and helium-depleted exterior are clearly seen in the function $c_0$ which tracks the contribution of radius shells to the planet's total mass (thin solid, black). Diamonds show the radius where half of the final $J_{\rm 2n}$ value is reached.}
\label{fig:contribs}
\end{figure}

In summary, assuming that the {\it Juno} and {\it Cassini} gravity experiments can successfully measure higher moments out to $\sim  J_{10}$ to a precision $\sim 10^{-8}$ or even
$\sim 10^{-9}$,
and assuming that the value of Saturn's $m$ can be accurately determined,
one can use TOF methods with adequate
accuracy to provide strong constraints on acceptable barotropes, particularly at radius levels relatively close to the surface, as shown in Figure \ref{fig:contribs}.  This in turn feeds back into the inferred core mass, even though the core region does not contribute directly to the multipole weighting functions.

\subsubsection{Adiabatic Assumption} \label{assumption}
The thermal structure of a planet interior is of importance in several ways. It will determine the phase of the material (liquid or solid) and it will likely play a role in its electronic properties (important for maintaining a magnetic field). It also contributes significantly to the pressure at a given density (of order 10\% is typical in the deep interior of Saturn.) It plays a central role in the thermal evolution of the planet; by the Virial theorem, the decrease of total heat content with time can be a large source of luminosity. It is also evident from the First Law that the planet is likely to be hot immediately after formation. One clearly needs a prescription for the thermal structure and how it evolves with time.

It is often said that the giant planets are ``adiabatic.'' There are good reasons for thinking that this is a good starting approximation but also good reasons for doubting that it is correct for the planets as a whole; here we review both perspectives. But first, a definition. The adiabatic assumption is more precisely stated as follows: Throughout most of the planet, the specific entropy within well-mixed layers is nearly constant. If there are several well-mixed layers (but with each layer having a different composition), then the entropy within each layer may also be nearly constant and the ``jump condition'' across layer interfaces is nearly isothermal. The assumption must of course break down in the outermost region of the atmosphere (optical depth unity or less), where outgoing IR photons are free to escape to space. But our concern here is whether and to what extent it breaks down deeper in the planet. 
Note that the preferred word is ``isentropic,'' not ``adiabatic'' since the latter word describes a process while the former is a thermodynamic statement, and that is what we need to define the thermal state. Nevertheless, we here apply the latter word in line with common usage.

It is instructive to first consider a completely homogeneous, fluid planet that is emitting energy from its interior (as Jupiter and Saturn are observed to do). As one proceeds deeper than optical depth unity, the opacity increases, primarily because of pressure induced molecular hydrogen opacity \cp{Guillot94a}.  As a consequence the temperature gradient that would be needed to carry out the heat by radiation alone exceeds the adiabatic temperature gradient, at least for most of the interior. As shown by \ct{Guillot94a}, a small radiative window may be possible at 2000-3000 K in Saturn. However, application of improved opacities for the alkali metals sodium and potassium leads to closure of such a window for Jupiter \cp{Guillot03}. The same is expected to hold for Saturn,
so that we safely assume that the opacities are too high for radiation to carry the heat efficiently throughout Saturn's interior.
 The heat carried by conduction throughout the interior once the hydrogen becomes an electrical conductor is also too small to be important. This means the interior is convectively unstable. Should such a planet be convecting heat outwards, then there is no doubt that the temperature gradient is extremely close to being adiabatic. The conclusion is reached in two steps: First, one asks what the state of the material would be if it were in fact isentropic. The answer is that it is everywhere fluid and of low viscosity. (In this context, even a viscosity six orders of magnitude greater than everyday water would qualify as ``low.'' In fact, the viscosity is comparable to that of everyday water). The second step is to recognize that convection in such a medium is extremely efficient in carrying heat over large distances. 

Consider, for example, the flows that could carry a few Watts m$^{-2}$ (typical of Saturn's interior ). We can write the heat flux as $F\sim \rho C_p v \delta T$, where $\rho$ is fluid density, $C_p$ is specific heat, $v$ is convective velocity and $\delta T$ is the temperature anomaly whose resulting density anomaly leads to the buoyancy that is responsible for $v$. We further expect $v \sim ( g \alpha \delta T L)^{1/2}$ where $\alpha$ is the coefficient of thermal expansion and $L$ is the characteristic length scale of the motions, because viscosity is too small to be relevant. For the choices $\rho$ =1000 kg/m$^{-3}$ , $C_p$ = $2\times10^4$ J kg$^{-1}$ K$^{-1}$,  $\delta T = 10^{-4}$ K, $\alpha=10^{-5}$ K$^{-1}$, $L=10^6$m, one finds $v\sim 0.1$ m s$^{-1}$ and $F\sim 10^2$ W/m$^{-2}$ . This crude order of magnitude argument (mixing length theory) may well be wrong by an order of magnitude or even several (perhaps the value of $L$ is smaller because of the Coriolis effect) but the conclusion is inescapable: Temperature deviations from an isentrope are expected to be extremely small. This implies something remarkable: Given the outer boundary condition (the specific entropy at the top of the convective zone), one can determine the temperature at the center of the planet \cp[e.g.][]{Hubbard73}. Adding an isolated core (i.e., a core that does not dissolve in the overlying material) to an otherwise homogeneous planet does not significantly change this story.  However, there are several respects in which this picture could be in error:
\begin{enumerate}
\item Although the planet was heated by accretion, it may have formed in such a way that the deep interior had lower entropy than the outer regions. Since gravitational energy release per unit mass increases as the planet grows, this could even be a likely outcome.
\item Although the planet may be fluid almost everywhere, that does not preclude first order phase transitions. The temperature structure may be altered by these transitions.
\item The core, if any, may be soluble in the overlying hydrogen. This creates compositional gradients (even if no gradient were present at the end of accretion).
\item The accretion process might create compositional gradients because of the imperfect mixing that arises when incoming planetesimals break up in the envelope.
\item The presence of H/He phase separation may act as a barrier to convection if the growing He droplets do not rain down instantaneously.
\item Condensation of ices and latent heat release in the weather layer can lead to molecular weight gradients and to deviation from adiabaticity due to the different temperature gradients along moist and dry adiabats.
\end{enumerate}

The first of these concerns is probably less important than the others. While it is indeed true that the specific entropy distribution may be initially stable (low entropy towards the center of the planet), the outer regions undergo entropy decrease with time, and this is rapid when the planet has a high effective temperature, with the result that the planet can naturally evolve towards an isentropic state.

The second concern has two important cases to consider. First there might be a first order phase transition from molecular to metallic hydrogen. In the fluid phase this is referred to as the Plasma Phase Transition (PPT). Many aspects of this transition are still uncertain, but there is currently no evidence that it persists to the high temperatures typical of interior models at the relevant pressure \cp{McMahon12}, suggesting a continuous transition in planets. Were such a transition to exist, it could lead to a stable interface between the two phases, with a entropy discontinuity across the interface \cp{SS77b}.  The more important case of relevance to Saturn (and apparently to some extent Jupiter as well) is the limited solubility of helium in hydrogen, which can impose a helium composition gradient, which is discussed in Section \ref{variants}.

\subsection{Rotation rate uncertainty} \label{rotate}  
The rotation period of a giant planet is a fundamental physical property that is used for constraining the internal structure and has implications for the dynamics of the planetary atmosphere. 
Saturn's rotation period is still not well constrained. \emph{Cassini} has confirmed a time-dependence in Saturn's auroral radio emission found from a comparison of \emph{Ulysses} data to the earlier \emph{Voyager} 1 and 2 spacecraft observations \cp{Galopeau00}. 
Because Saturn's magnetic pole is aligned with Saturn's rotation axis, 
Saturn's standard rotation period is set to the \emph{Voyager 2} radio period, 10h 39m 22.4s. This rotation period was derived from the periodicity in Saturn's kilometric radiation SKR \cp[e.g.,][]{Dessler83}. Surprisingly, \emph{Cassini}'s SKR measured a rotation period of 10h~47m~6s \cp[e.g.,][]{Gurnett07}, about eight minutes longer using the exact same method. 
It is now accepted that Saturn's exact rotation period is unknown to within  several minutes and cannot be inferred from SKR measurements.   
In addition, atmospheric features such as clouds cannot directly be used to derive Saturn's rotation period because it is unclear how they are linked to the rotation of the deep interior, and in fact Saturn's observed wind velocities are always given relative to an {\it assumed} solid-body rotation period.

Recently, several theoretical approaches to determine Saturn's rotation period have been presented. The first approach was based on minimizing 
of the dynamical heights at the 100~mbar pressure-level above the geopotential surface caused by the atmospheric winds \cp{Anderson07}. The dynamical heights (as well as the wind speeds) were found to be minimized for a rotation period of 10h 32m 35s $\pm$ 13s, about 7 minutes
shorter than the \emph{Voyager 2} value. In a second study, an analysis of the
potential vorticity based on considerations of their dynamic meteorology was presented. Saturn's derived rotation
period was found to be 10h~34m~13s $\pm20$s \cp{Read09}. Both of these studies relied on the measured wind velocities obtained from cloud tracking at the observed cloud level.

In a third study, Saturn's rotation period was derived from its observed gravitational moments and its observed shape including uncertainties in these measurements by taking an optimization approach \cp{Helled2015}. The gravitational data are insufficient to uniquely determine the rotation period, and therefore, the problem is under-determined (there are more unknowns than constraints). Accordingly, a statistical optimization approach was taken, and by using the constraints on the radius and the gravitational field, the rotation period of Saturn was determined (statistically) with a fairly small uncertainty. 
When only the gravitational field is used as a constraint, the rotation period was found to be 10h 43m 10s $\pm4$m. With the constraints on Saturn's shape and internal density structure, the rotation period was found to be 10h 32m 45s $\pm 46$s, in excellent agreement with \ct{Anderson07}. This is because Saturn's mean radius is more consistent with shorter rotation periods, {\it if} dynamical distortions on the shape are not included. 
Interestingly, all of these studies infer a shorter rotation period for Saturn than the \emph{Voyager 2} rotation period, leading to smaller wind velocities and atmosphere dynamics more similar to that of Jupiter. The validity of these theoretical approaches, however, is yet to be proven and a more compete understanding of the shape-dynamics-internal rotation feedback is required.  However, all three of these methods described above, when applied to Jupiter, yield a  rotation period that is consistent with Jupiter's generally accepted value, based on the rotation of its non-axisymmetric magnetic field.

\begin{figure*}[h!]
\centering
\includegraphics[width = 0.8\textwidth]{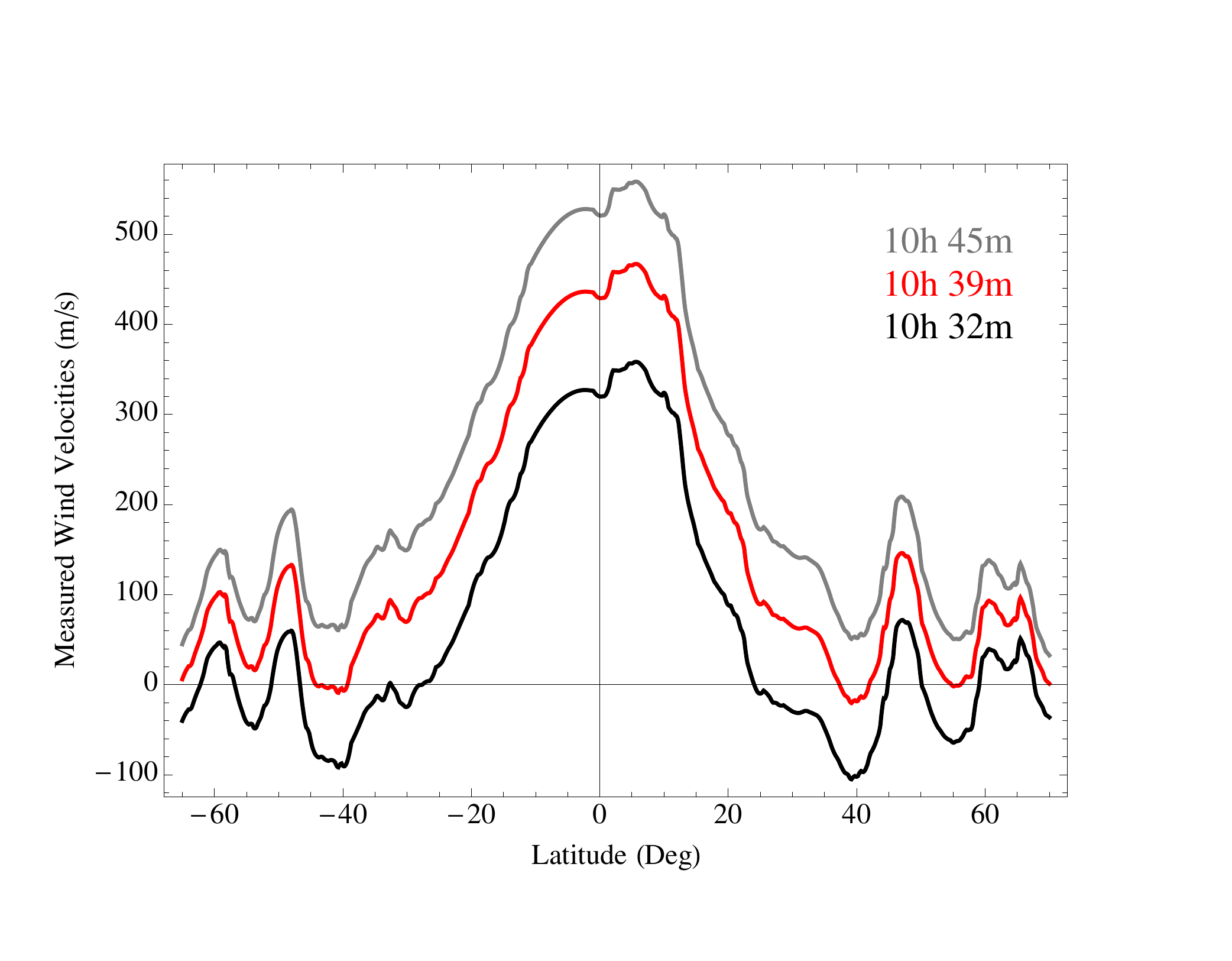}
\vspace{-35pt}
\caption{Saturn's wind velocities for three different underlying rotation periods: 10h 32m (black), 10h 39m (red), 10h 45m (gray). The measured wind velocities at the cloud-layer are obtained from \ct{Sanchez00}.}
\label{fig:wind-sat}
\end{figure*}

Besides implications of the rotation period of Saturn to its inferred internal structure (see below), it also directly affects the atmospheric wind velocities. 
Saturn's wind profile with respect to three different rotation periods is shown in Figure \ref{fig:wind-sat}. Shown are the wind velocities vs.~latitude (degrees) for rotation periods of 10h 32m (red), 10h 39m (black), and 10h 45m (gray). A rotation period of about 10h 32m implies that the latitudinal wind structure is more symmetric, containing both easterly and westerly jets as observed on Jupiter.

Finally, an additional complication regarding Saturn's internal rotation period arises from the fact that Saturn could rotate differentially on cylinders and/or that its atmospheric winds penetrate deep into the interior. 
This can also affect interior models \cp[e.g.,][]{Hubbard82, Hubbard99b, Helled13}. 
The realization that Saturn's rotation period is not constrained within
a few percent, and the possibility of differential rotation, introduce an uncertainty for interior models as discussed below. 
\par

\subsection{Current structure from isentropic models}\label{sec:struc_hom}

Saturn's internal structure has been studied for decades. In this Section, we concentrate on results from recent interior models of Saturn, and on the physical parameters that are used to constrain the planetary interior. 

\subsubsection{Observational constraints and model assumptions}  \label{sec:models_merged}

Constraints on Saturn's internal structure have been provided through several spacecraft missions and ground-based observations.  For brevity, we call such constraints ``observational'' although most of them are not directly obtained from measurements but through a variety of theoretical models that fit the measured data. 
Prior to the \emph{Cassini} era, constraints were derived for Saturn's total mass, shape, gravitational harmonics ($J_{2n}$), periodicities in its surrounding plasma disk, magnetic field, and kilometric radio-emission, the atmospheric helium abundance and temperature profile, and the luminosity.  \emph{Cassini} has improved our precision on most of these quantities.

The uncertainty in Saturn's internal structure is not only linked to the EOS and the uncertainties in the observational constraints, but is in fact, in the philosophy adopted when modeling the interior. Even under the assumption of an isentropic interior, model properties such as the existence of differential rotation, the number of layers, and the distribution of heavy elements, can lead to rather different inferred structures and compositions. 
For example, the consideration of the expected correction of differential rotation to the gravitational field calculated by static models relaxes the otherwise rather stringent 
constraints of the measured gravitational field assuming solid-body rotation. In fact , there are various estimates for the magnitude of this effect \cp[e.g.,][and chapter on Saturn's atmosphere dynamics by Showman et al.]{Hubbard1982,Kaspi2013} and they are crucial for the interpretation of Cassini gravity data. 
It should be noted however, that even for a perfectly known gravitational field and the contribution of dynamics (differential rotation) there is still ambiguity regarding the internal density distribution. This is reflected in the various assumptions adopted by different authors, such as inhomogeneities in heavy elements and the location of the transition from helium-poor to helium-rich envelopes.  
The internal density distribution can be affected by various physical processes such as helium rain and core erosion that are not well understood. 
As a consequence, there is some freedom in constructing an interior model - some authors include corrections due to differential rotation, while others add an inhomogeneity in heavy elements between the inner and outer H/He envelope (e.g., $Z_{\rm out} \neq Z_{\rm in}$ - see below). Standard interior models assume a three-layer structure, since the goal is to minimize the number of free parameters in the models.

The uncertainties in Saturn's $J_2$, $J_4$, $J_6$ have been significantly reduced compared to the pre-\emph{Cassini} era through the combined analysis of \emph{Pioneer 11}, \emph{Voyager}, \emph{Cassini}, and long-term ground-based and \emph{HST} astrometry data \cp{Jacobson06}. Smaller error bars provide an opportunity to narrow down the possible internal density distributions and thus Saturn's internal structure. 
Table \ref{tab:nn_obs} lists some of the values applied to Saturn interior models. As discussed above, uncertainties in Saturn's rotation rate and the fact that atmospheric winds and/or differential rotation affect the planetary shape cause an additional uncertainty in values of the gravitational harmonics used by hydrostatic interior models. Several models  have accounted for this uncertainty as presented below.

\begin{table*}
\begin{tabular}{lllllllll}\hline
Ref. & $R_{\rm eq}$ (km) & $J_2\times 10^2$ & $J_4\times 10^4$ & $J_6\times 10^4$ & $T_{1-\rm{bar}}$ (K) & $Y_{\rm {atm}}$ & $Z_{\rm env}$ & Diff.~Rot.  \\\hline
\multicolumn{6}{l}{\bf{Voyager constraints}}\\
 G99,   & 60268(4) & 1.63320(100)$^{\ast}$ & -9.190(400)$^{\ast}$ & 1.04(50)$^{\ast}$ & 130-140 & 0.11-0.21 & inhom  & Yes\\
SG04  & 60268(4) & 1.63320(100)$^{\ast}$ & -9.190(400)$^{\ast}$ & 1.04(50)$^{\ast}$ & 135-145 & 0.11-0.21 &  hom  & Yes\\
 HG13 &  60269 & 1.62580(410) & -9.050(410) & 98(51) & 130-145 & 0.11-0.25 & hom  & Yes \\
\multicolumn{6}{l}{\bf{Cassini constraints}}\\
 H11 &  60141.4 & 1.63931$^{\ast}$ & -9.476$^{\ast}$ & 87.8$^{\ast}$ & $\approx$ 135 & --- & inhom & No\\
 N13 &  60268 & 1.63242(3)$^{\ast}$ & -9.396(28)$^{\ast}$ & 86.6(9.6)$^{\ast}$ & 140  & 0.18 & inhom & No\\
 HG13 & 60269 & 1.62510(400) & -9.260(110) & 81(11)  & 130-145 & 0.11-0.25 & hom & Yes\\ 
\hline
\end{tabular}
\caption{\label{tab:nn_obs} Values for the gravitational coefficients applied to models of Saturn with a nominal rotation period of 10h 39m 24s.  Saturn's measured gravitational coefficients are determined to be $J_2\times 10^2$ = 1.62907(3); $J_4\times 10^4$ = -9.358(28) $J_6\times 10^4$ = 86.1(9.6) for a reference radius of 60,330 km \cp{Jacobson06}. $R_{\rm eq}$ corresponds to the equatorial radius at the 1-bar pressure level. $T_{1-\rm{bar}}$ is the assumed temperature at 1 bar, $Y_{\rm atm}$ is the atmospheric helium mass fraction and $Z_{\rm env}$ being the envelope metallicity which is either assumed to be homogeneous (hom) or inhomogeneous (inhom), and Diff.~Rot.~corresponds to whether corrections linked to differential rotation were considered (Yes/No). G99: \cite{Guillot99}; HG13: \cite{Helled13}; H11: \cite{Helled11a}, J06: \cite{Jacobson06}; N13: \cite{Nettelmann13b}; SG04: \cite{Saumon04}}
\end{table*}


We first describe the different model assumptions and imposed constraints of recent isentropic, quasi-homogeneous Saturn models \citep{Guillot99,Saumon04,Helled13,Nettelmann13b}. 

\begin{figure}[h!]
\centering
\includegraphics[width=0.5\textwidth]{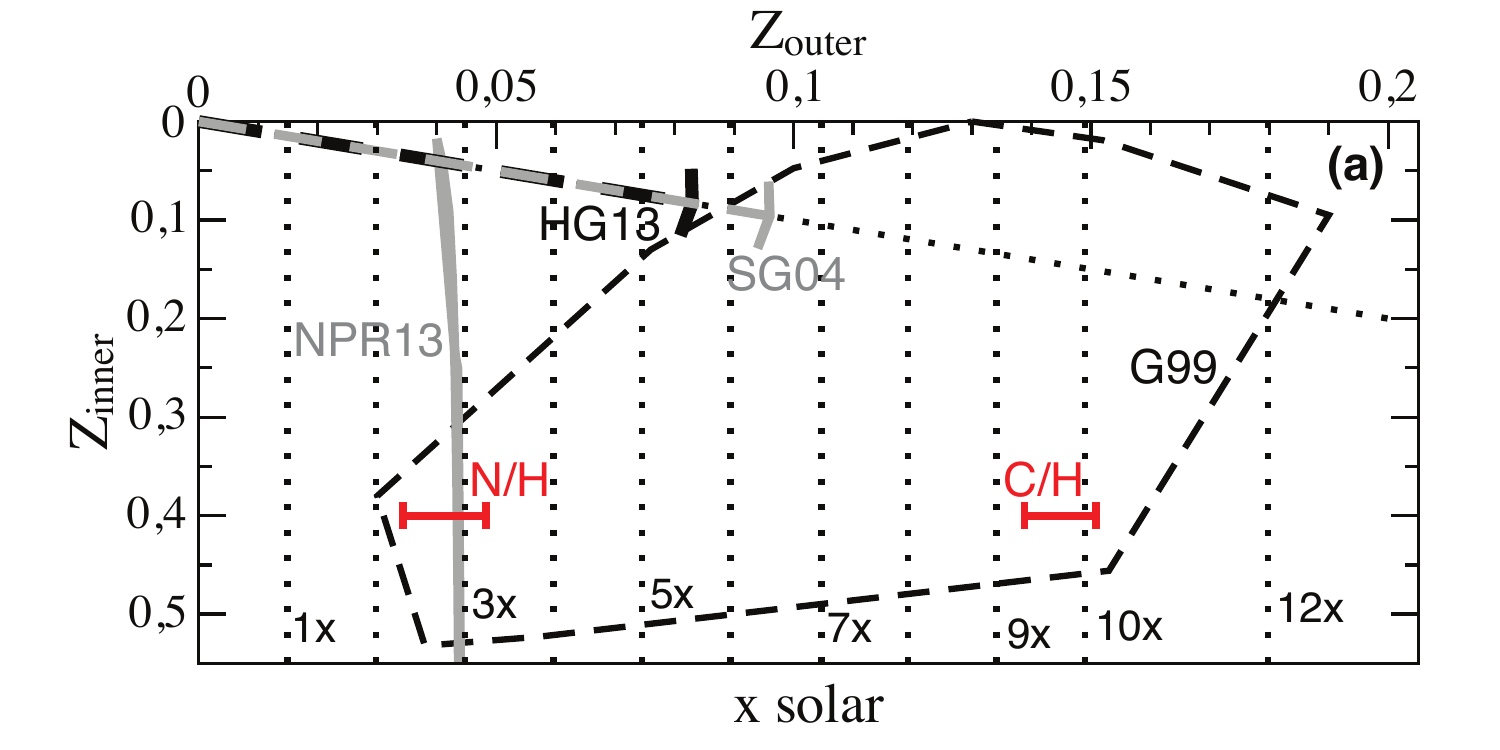}
\includegraphics[width=0.5\textwidth]{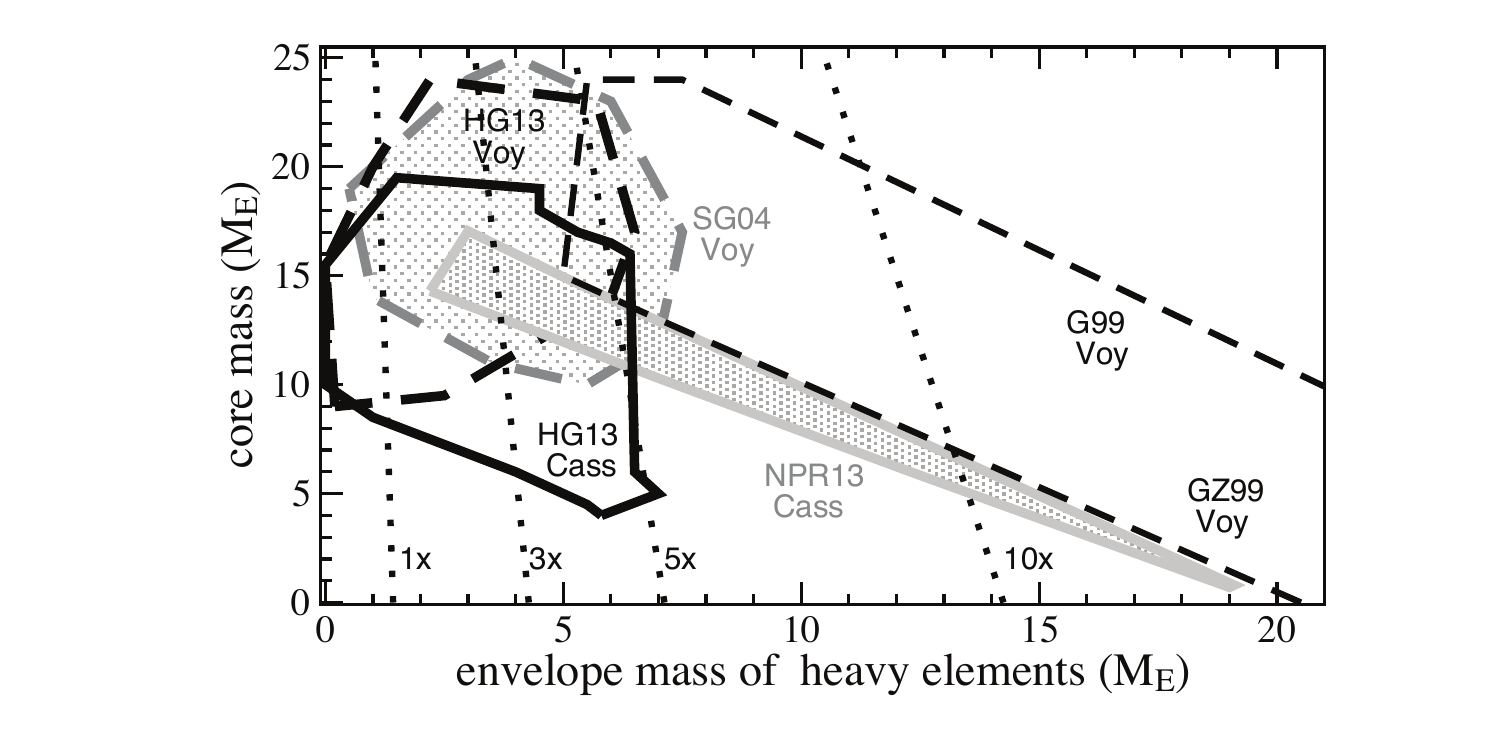}
\caption{\label{fig:nn_McZZ} Saturn structure model results under various assumptions made by different authors, see text for details. (a) heavy element enrichment by mass in the outer and inner envelope; enrichment factors over the bulk protosolar abundance of $Z_0=0.0149$ refer to $Z_{outer}$ and are compared to the measured C/H and N/H ratios in Saturn's atmosphere. (b) core mass and total mass of heavy elements in the homogeneous, isentropic envelopes. }
\end{figure}

Saturn models by \ct{Guillot99} (enclosed by \emph{thin black dashed lines} in Figure \ref{fig:nn_McZZ}) were designed for consistency with the \emph{Voyager} constraints.  A surface temperature of $T_{\rm 1 bar}=135-145$~K, a rotation period of 10h 39m, a bulk helium mass fraction of $Y=0.265-0.285$ with an atmospheric helium mass fraction of $Y_{\rm atm}=0.11-0.21$ were used. The interior models were derived using the SCvH EOS for He and the SCvH-i EOS for hydrogen \cp{SCVH}, which  interpolates between the EOSs for the molecular and the metallic phases of hydrogen. Furthermore, they assume a sharp layer boundary between the molecular and metallic hydrogen envelope, where the abundances of helium and of heavy elements ($Z_{out}$ for heavy elements in the outer envelope and $Z_{in}$ for heavy elements in the inner envelope, respectively) change discontinuously. Heavy elements were assumed to have a water-like mean molecular weight and specific heat. 
Their mass fraction is derived from any ``excess'' helium abundance needed to fit the \emph{Voyager} constraints.

\cite{Helled13} modeled Saturn's interior assuming a three-layer structure that consists of a central ice/rock core and an envelope that is split into a helium-rich metallic hydrogen region and a helium-poor molecular region.  The main differences in underlying model assumptions from those of \cite{Guillot99} consist in $Y_{\rm atm}=0.11-0.25$, a global $Y=0.265-0.275$ consistent with the protosolar value \cp[e.g.][]{Bahcall95}, $T_{\rm 1 bar}=130-145$ K, and, perhaps most importantly, in imposing $Z_{in}=Z_{out}$ and using a physical EOS of water and sand for heavy elements in the envelopes. 
The transition pressure between the helium-rich to helium-poor (hereafter $P_{{\rm{trans}}}$) was assumed to be between 1 and 4 Mbars. The envelope heavy elements were assumed to be homogeneously mixed within the planetary envelope, as may well be expected in models that lack a first-order phase transition for hydrogen.
In order to account for the uncertainty associated with differential rotation, the uncertainty in the gravitational harmonics as expected from differential rotation was also included \cp[e.g.][]{Hubbard1982}. In addition, two sets of gravitational data were used: the gravitational moments as measured by \emph{Voyager} and those measured by \emph{Cassini} (see Table 3).  Two solid-body rotation periods were considered: the \emph{Voyager 2} radio period, and a shorter period that was set to the 10h 32m 35s as derived by \ct{Anderson07}. Finally, to account for the uncertainty in Saturn's shape three different cases in terms of mean radius were considered. $c_0$:  the ``standard'' equatorial radius of 60,269 km as previously assumed by interior models.  $c_1$:  a fixed polar radius with a corresponding equatorial radius of 60,148 km. $c_2$: an intermediate case - with an equatorial radius set to 60,238 km. 
Figures \ref{fig:HG13}a,b show the resulting models, while Figure \ref{fig:nn_McZZ} shows subsets for the \emph{Voyager} rotation rate and respectively the \emph{Voyager} constraints (within the \emph{thick black dashed lined}) and the \emph{Cassini} constraints (within the \emph{black solid lined area}.

Models by \cite{Saumon04} (\emph{grey dashed lined} in Figure \ref{fig:nn_McZZ}) use the same observational constraints as in \cite{Guillot99} but assume $Z_{in}=Z_{out}$ as \cite{Helled13}; $P_{trans}$ between helium-poor and helium-rich varies from 1 to 3 Mbar.
The \emph{grey shaded areas} in Figure \ref{fig:nn_McZZ} shows interior models derived by \citet{Nettelmann13b}, based on \emph{ab initio} EOS for hydrogen, helium, and H$_2$O, with $T_{\rm 1 bar}=140$~K, $Y_{\rm atm}=0.18$, and $Y=0.275$, an allowance for $Z_{in}\not= Z_{out}$, cores made of pure rock or pure water, and no imposed limit on $P_{trans}$ .
Finally, the models by \citet{Gudkova99} resemble those of \cite{Guillot99} but assume a five-layer structure with a helium layer on top the core to formally account for hydrogen-helium demixing and helium sedimentation in the entire inner envelope.

\subsubsection{Results for isentropic models}

\begin{figure*}[h!]
\centering
\vspace{-45pt}
\includegraphics[width = 0.49\textwidth]{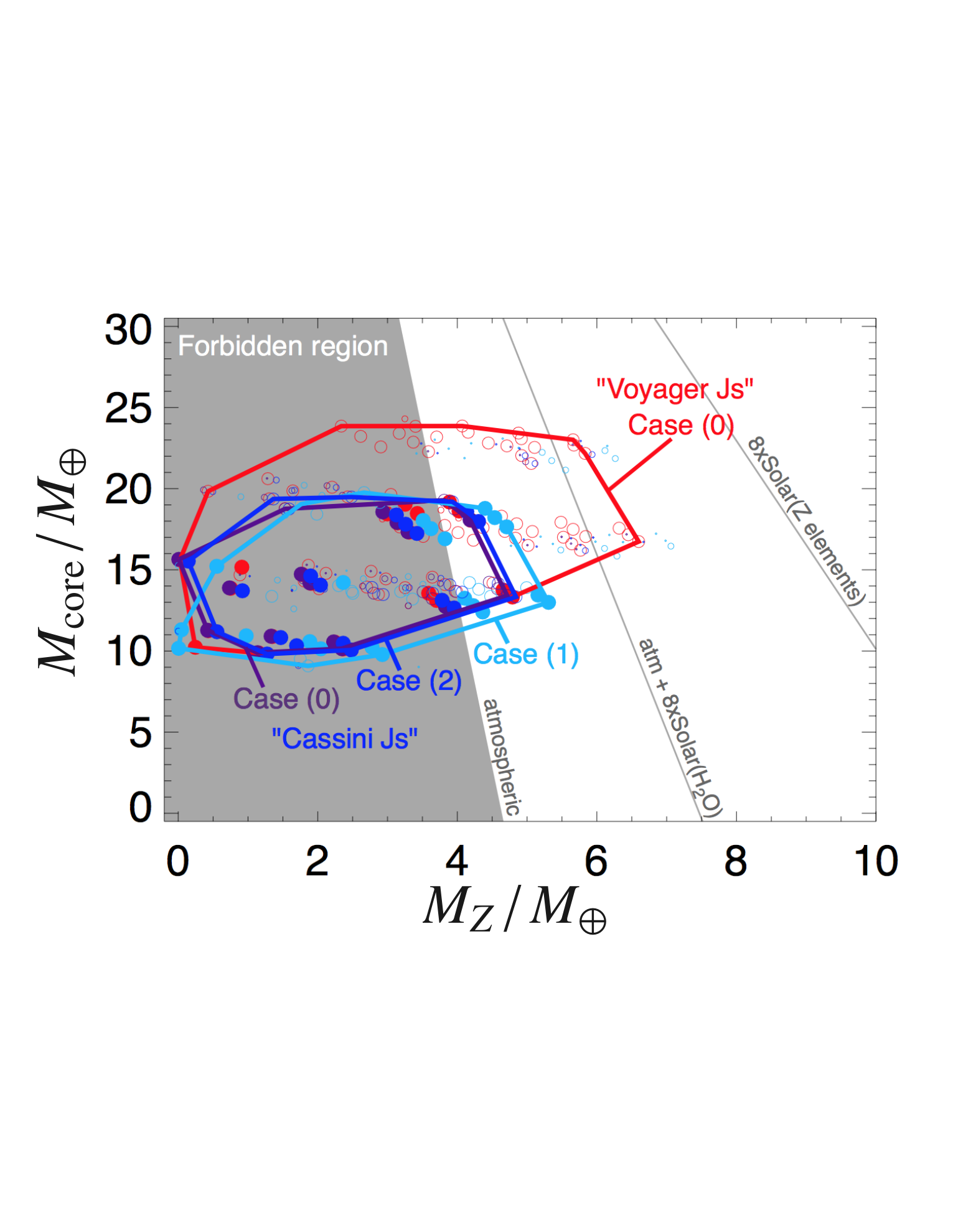}
\includegraphics[width = 0.49\textwidth]{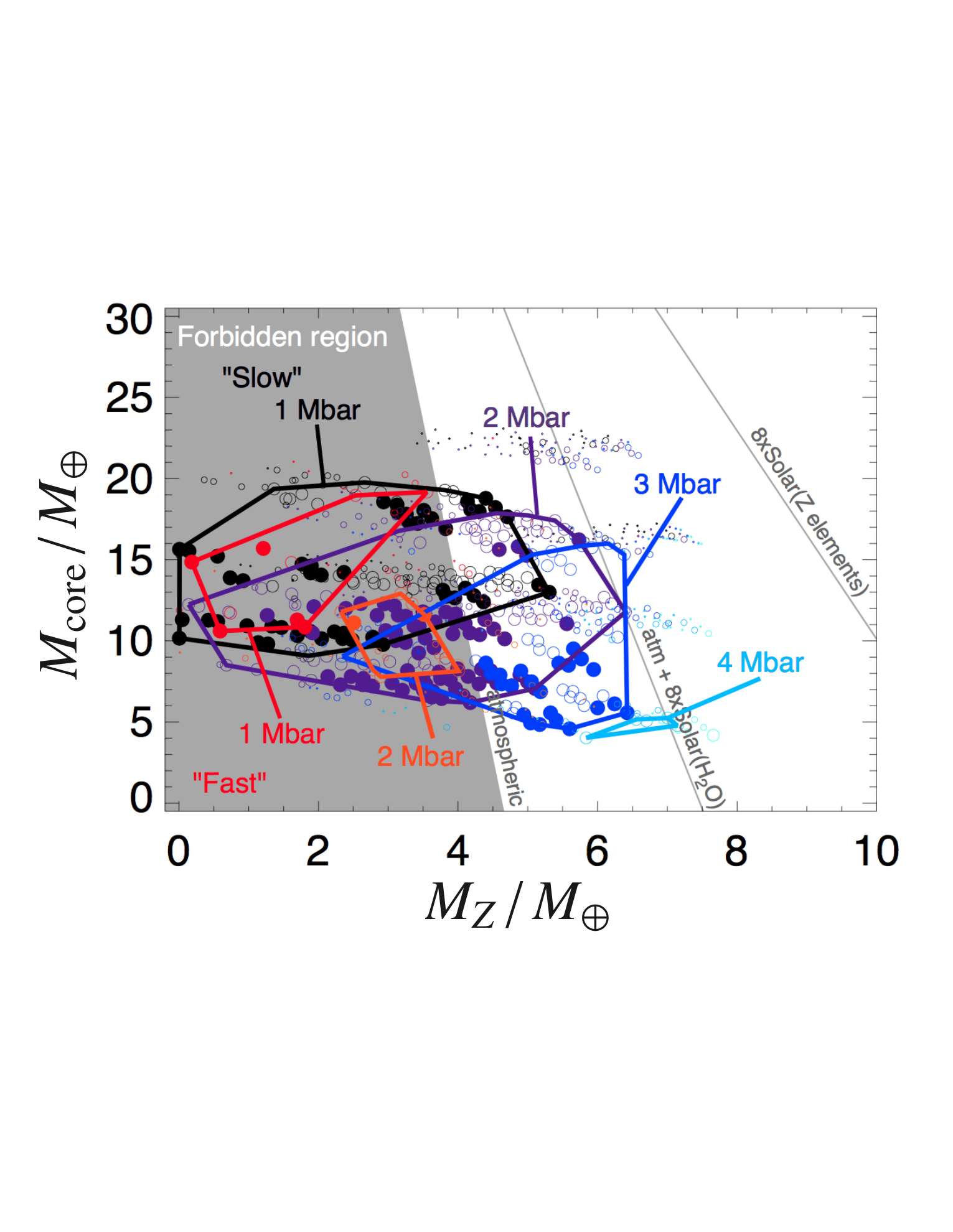}
\vspace{-70pt}
\caption{Saturn's core mass (M$_{\rm{core}}$) vs.~the mass of heavy elements in the envelope (M$_{\rm Z}$) for interior models matching the available observational constraints. {\bf Left:} Solutions for $P_{{\rm{trans}}}$ = 1 Mbar using the Voyager rotation period with \emph{Voyager}'s $J$s, and model $c_0$  (red), and for \cas\ $J$s and models $c_0$ (purple), $c_1$ (blue), $c_2$ (light blue). {\bf Right:} Solutions when using the \cas\ $J$s, combining three different cases for the planetary shape ($c_0, c_1, c_2$): (i) \emph{Voyager} rotation period and $P_{{\rm{trans}}}$ = 1 Mbar (black), 2 Mbar (purple), 3 Mbar (blue), 4 Mbar (light blue). (ii) A rotation period of 10h 32m 35s and $P_{{\rm{trans}}}$ = 1 Mbar (red), 2 Mbar (orange). From Helled and Guillot (2013).}
\label{fig:HG13}
\end{figure*}

Figure \ref{fig:nn_McZZ}b presents the derived masses of heavy elements in the core, and in the envelope, $M_{Z,env}$. Figure \ref{fig:nn_McZZ}a presents the envelope heavy element mass fractions for isentropic Saturn models obtained under various model assumptions and by different authors as described in Section \ref{sec:models_merged}.  The sensitivity of the derived internal structure to the assumed shape and rotation rate is demonstrated in Figure \ref{fig:HG13}a,b where we show the results by Helled \& Guillot (2013) when assuming various rotation periods and shapes. 

From Figure \ref{fig:nn_McZZ} several striking properties can be seen: 
$(i)$ the \emph{Cassini} gravitational data reduces Saturn's core mass by $\approx 5$ \me, i.e.~from $\sim 10-25$ \me\ to $\sim 5-20$ \me, $(ii)$ models that allow for a larger enrichment in the deep envelope than in the atmosphere ($Z_{\rm out} < Z_{\rm in}$) allow for no-core solutions if $P_{trans}\approx 5$ Mbar. 
$(iii)$ the models differ largely in their predicted heavy element mass fraction and $Z_{\rm out}$ values, which for the \emph{Cassini} data is found to range from 0.1 to $5\times$ solar \citep{Helled13}, or from 2 to $13\times$ solar \citep{Nettelmann13b}, the difference is mostly due to the EOS of heavy elements in the envelopes (water+sand vs.~pure water), and due to the $J_2$ and $J_4$ values used. 

Since the heavy element mass fraction is unlikely to decrease with depth, as it would trigger an instability that would tend to equilibrate the abundances, it is reasonable to assume $Z_{atm} \leq Z_{out}$ (and potentially that $Z_{atm} \sim Z_{out}$) and $Z_{out} \leq Z_{in}$. Therefore, an observational determination of the bulk atmospheric heavy element abundance $Z_{\rm atm}$ through measured O/H, C/H, and N/H ratios below the respective cloud decks can be used to rule out a vast amount of Saturn  models. This idea is highlighted in Figure \ref{fig:nn_McZZ}a, where measured C and N enrichments \citep{Guillot14} are plotted in comparison to $Z_{out}$ values from the structure models. Models based on the \emph{Cassini} constraints \citep{Nettelmann13b,Helled13} are consistent with the measured $\sim 3\times$ solar enrichment of N/H, which, however, may only be a lower limit to the abundance at greater depths, whereas only the early models of \cite{Guillot99} seem to allow for bulk $9\times$ solar enrichments as indicated by the C/H ratio. We conclude that modern Saturn models based on tighter constraints for the gravitational field and from updated EOS calculations predict $Z_{\rm atm}$ less than $\sim 5\times$ solar, and thus O/H less than $\sim 7\times$ solar \citep{Nettelmann13b}.

The results for Saturn's internal structure as derived by Helled \& Guillot (2013) are shown in detail in Figure \ref{fig:HG13}. The left panel presents a comparison of the derived interior model solutions for Saturn with $P_{{\rm{trans}}}$ = 1 Mbar, for the \emph{Voyager} rotation period and with \emph{Voyager} $J$s, and model $c_0$ (red), and for \emph{Cassini} $J$s and models $c_0$ (purple), $c_1$ (blue), $c_2$ (light blue). The contours correspond to interior models that fit within 2 sigma in equatorial radius, $J_2$, and $J_4$ ($J_6$ fits within 1 sigma). The grey area represents a ``forbidden zone''  corresponding to a region that its atmospheric abundances are inconsistent with  the atmospheric abundances derived from measurements \cp[e.g.][]{Guillot14}. The first grey line adds 8 times the solar abundance \cp{Asplund09} of water, while the second grey line assumes that all the heavy elements are enriched by a factor of 8 compared to solar. 
It is found that the possible parameter-space of solutions is smaller when \emph{Cassini}'s $J$s are used. This is not surprising given that the uncertainties of the gravitational harmonics are smaller. 
With the \emph{Voyager} rotation period the inferred heavy element mass in Saturn's envelope is 0 -- 7 M$_{\oplus}$ and the core mass is 10--20 M$_{\oplus}$. 
The right panel of Figure \ref{fig:HG13} shows how the assumed rotation period affects the inferred composition of Saturn. 
For the \emph{Voyager} period Saturn's core mass ranges between 5 and 20 M$_{\oplus}$ with the lower values corresponding to higher $P_{{\rm{trans}}}$. Saturn's core mass strongly depends on the $P_{{\rm{trans}}}$ where the derived core mass decreases significantly for higher transition pressures.  
The heavy element mass in the envelope remains 0-7 M$_{\oplus}$. Interior models with the shorter rotation period with $P_{{\rm{trans}}}$ = 1 and 2 Mbar were found to have heavy element mass in the envelope less than 4 M$_{\oplus}$, below the values derived from atmospheric spectroscopic measurements. Solutions with this rotation period can be found when a discontinuity in the heavy elements distribution is considered (see e.g., Nettelmann et al., 2013). It should also be noted that {\it interior models of Saturn with no ice/rock core are possible}. 
The lack of knowledge on the depth of differential rotation in Saturn, its rotation period, and whether the heavy elements are homogeneously distributed within the planet are major sources of uncertainty on the internal structure and global composition of the planet.

Despite the uncertainties in Saturn's derived internal structure, an important conclusion from these results can be drawn: none of the Saturn models has a $P_{trans}$ value near the core-mantle boundary of 10--15 Mbar, although a wide range of uncertainties has been considered. Hence, an interior where helium rains down all the way to the core and leaves the envelope above homogeneous and isentropic, is \emph{excluded}.  In other words, if helium in Saturn rains down to the core, the mantle above should have an 
inhomogeneity in helium (or heavy elements). This inhomogeneous region could be non-isentropic. This conclusion holds unless the helium layer on top of the core reaches upward to $\sim 4$ Mbar, in which case Saturn's atmosphere should be highly depleted in helium, or unless the gravitational harmonics are severely altered by deep winds. A measurement of high-order gravitational harmonics and of the atmospheric helium abundance are thus important for establishing a better understanding of Saturn's internal structure.

\subsubsection{Comparison with Jupiter} 
It is useful to compare and contrast Jupiter and Saturn.  While both Jupiter and Saturn are massive giant planets that consist mainly of hydrogen and helium, their relative enrichment compared to protosolar composition is somewhat different, and according to most of the available interior models, Saturn is predicted to be more enriched with heavy elements. 
In addition, while interior models for both planets suggest that solutions with no cores are valid, typically, Saturn's interior models include a core, and indeed the existence of a dense inner region in Saturn is  supported by its lower MOI value.  
Both planets are fast rotators, having a large equatorial jet, but without constraining the rotation period of Saturn exactly, we cannot say whether both planets have eastern and western jets at high latitudes or whether this feature exists only for Jupiter, and how the two planets compare in terms of wind speeds.
In addition, while the atmospheres of both planets are depleted in helium compared to protosolar composition, the depletion appears to be more significant for Saturn. As discussed below, this leads to the ``slower" cooling of Saturn and its luminosity excess. Finally, other important differences are the heat flux, tilt, strength of the magnetic fields, and the rings and satellites systems.

Is Saturn simply a smaller version of Jupiter? Not necessarily. Given our current understanding of planet formation, the cause of the difference in terms of total mass and composition seems to be  the slower formation of Saturn compared to Jupiter. In the standard picture of giant planet formation, also known as core accretion \cp[see e.g.,][and the Atreya et al. chapter on Saturn's origin]{Helled14c}, the growth rate of a planetary embryo is larger at small radial distances, which explains why Jupiter could have reached a critical mass for runaway gas accretion before Saturn. Thus, both planets have reached critical masses and accreted significant amounts of hydrogen and helium. Why is Saturn's gaseous envelope smaller? This is not yet fully understood, but is most likely related to Saturn's relatively slower growth rate and the interaction with Jupiter, and possibility, other growing planets (i.e., Uranus and Neptune). This suggestion has been investigated in several versions of the Nice model \cp[e.g.,][and references therein]{Thommes1999,Tsigani2005,Walsh11}. 

There are still many open questions regarding the origin and internal structure of Jupiter and Saturn and it is certainly beneficial to study the two planets together. An opportunity to investigate and compare Jupiter and Saturn will be possible in the upcoming years. By 2017, accurate measurements of the gravitational fields of the planets will be available from the \emph{Juno} and \emph{Cassini Solstice} missions. A detailed comparison of Jupiter and Saturn will be very inspiring and improve our understanding of the origin of the solar-system and will provide insights on the characteristics of gas giant planets in general.

\subsection{Results from Thermal Evolution Modeling} \label{evol}
\subsubsection{Simple models with helium phase separation}
\begin{figure}[h!]
\includegraphics[width = 0.5\textwidth]{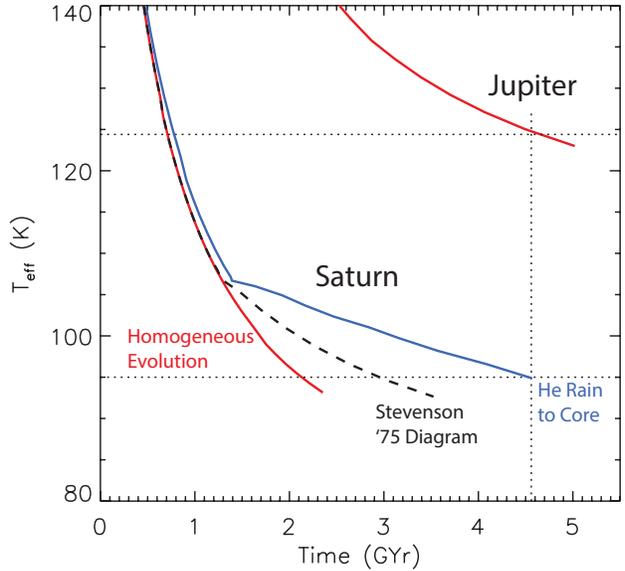}
\caption{Fully isentropic, homogeneous models of the thermal evolution of Jupiter and Saturn (red), after \ct{FH03}.  The current \teff\ of each planet is shown with a dotted line.  For Saturn, the real planet (age 4.55 Gyr) has a much higher \teff\ than the model, indicating the model is missing significant physics.  A model including helium-rain using the \ct{Stevenson75} phase diagram is shown in dashed black.  A model that uses an ad-hoc phase diagram, designed to rain helium down to the top of the planet's core, liberating more gravitational energy, is shown in blue.}\label{js}
\end{figure}

The first thermal evolution calculations for warm, fluid, adiabatic models of Jupiter and Saturn were performed by a number of authors in the mid 1970s \cp{Graboske75,Bodenheimer76,Hubbard77,Pollack77} and most of their findings remain relevant today.  Models for Jupiter, starting from a hot post-formation state with a hydrogen-helium envelope that was assumed homogeneous, isentropic, and well-mixed, cooled to Jupiter's known \teff\ of 124 K in $\sim$4.5 Gyr.  However, these calculations failed to reproduce Saturn's cooling history.  A Saturnian cooling age of 2-2.5 Gyr was found, implying that Saturn today (\teff=95 K) is much too hot, by a \emph{factor of 50\% in luminosity} \cp{Pollack77,SS77b}.  (See Figure \ref{js}.)  If giant planets are fully or mostly isentropic below their radiative atmospheres, then it is the atmosphere that serves as the bottleneck for interior cooling.  However, advances in atmosphere modeling used in several generations of thermal evolution models \cp{Graboske75, Hubbard99, Fortney11} did not alter this dichotomy in the cooling history between Jupiter and Saturn.
The leading explanation to remedy the cooling shortfall has long focused on the separation of helium from hydrogen (or ``demixing'') \cp[e.g.][]{SS77b,SS77a,FH03}.  Stevenson \& Salpeter found that a ``rain'' of helium was likely within Saturn, and perhaps Jupiter, and that this differentiation (a conversion of gravitational potential energy to thermal energy) could prolong Saturn's evolution, keeping it warmer, longer.  (See Figure \ref{js}.)  The evidence for this process is strong if Jupiter and Saturn are considered together.  The atmospheres of both Jupiter and Saturn are depleted in helium, relative to the protosolar abundance (mass fraction $Y_{\rm proto}=0.270 \pm 0.005$, derived from helioseismology \cp{Asplund09}).  For Jupiter, $Y_{\rm atmos}=0.234 \pm 0.008$, from the \emph{Galileo Entry Probe} \cp{vonzahn98}.  Furthermore, Ne is strongly depleted in the atmosphere as well, and calculations suggest it is lost into He-rich droplets \cp{Roulston95,Wilson10}.  Taken together, this is convincing evidence for phase separation.  One would expect that Saturn, being cooler than Jupiter, should be more depleted in He. However, \emph{Voyager 2} estimates from spectroscopy run from $Y_{\rm atmos}=0.01-0.11$ \cp{Conrath84} to $Y_{\rm atmos}=0.18-0.25$ \cp[see][for details]{CG00}.  Revised estimates from \emph{Cassini} have not yet been published, but preliminary values cluster around $Y_{\rm atmos}\sim0.14$ (P.~Gierasch, pers communication).

These depletions imply that helium phase separation has begun in Jupiter, perhaps relatively recently, and that it has been ongoing in Saturn for a longer time.  There is no other published explanation for the planets' He depletions.  \ct{Hubbard99} and \ct{FH03} found that by raining He down all the way to the core at a late evolution state, that the maximum allowed $Y_{\rm atmos}$ must be $<0.20$ for Saturn today, to explain its current luminosity.  However, our understanding of Saturn is clearly not complete as the correct amount of He depletion, along with its distribution within the interior, must be accounted for in models.

Barriers to a better quantitative understanding of the phase separation process include:  1] The H/He phase diagram is still not precisely known, which dramatically impacts the amount of Saturn's mass within the He-immisciblity region, as well as the extent of any He-gradient region.  2] The issue of how He composition gradients would affect the temperature gradient in Saturn's deep interior is poorly understood, but new work in this area is described below.

\subsubsection{Variants for inhomogeneous structure and evolution models} \label{variants}
Homogeneous models of Saturn are successful in providing a good match to many observed properties 
such as the low-order gravitational harmonics. However, they fail to explain Saturn's high luminosity, 
its dipolar magnetic field (see Section \ref{sec:mag}) and the fine-spitting of density waves in its rings (see Section \ref{sec:seis}).
Saturn has therefore been suggested to contain at least one inhomogeneous zone in its interior. 
We review recent attempts of inhomogeneous model developments for Saturn and discuss their physical 
justifications as well as their abilities in explaining the observational constraints.

As discussed in Section \ref{assumption} an inhomogeneous zone in a giant planet can have different possible origins, due to the formation process itself, 
subsequent erosion of an initially massive core, or phase separation and sedimentation of abundant constituents.
While details of each of these processes are not well understood yet, some basic properties can naturally lead to
deviations from homogeneity. For instance, the core accretion scenario for giant planets suggests that 
during the period of massive core formation, before rapid run-away gas-accretion sets in, both gaseous material and 
planetesimals of various sizes are accreted. While small planetesimals may easily dissolve in the gaseous component, 
and heavy ones sink down to the initial core, medium-sized bodies may dissolve at different altitudes and cause a compositional gradient. Moreover, during the subsequent long-term evolution, the initial massive core may then erode as the heavy elements (O, Si, Fe) are miscible in the metallic 
hydrogen  envelope above \citep{Wilson12a,Wilson12b,Wahl13}. The efficiency of upward mixing by thermal convection may be low 
and take billions of years,  so that today an inhomogeneous region may still exist atop the core
\citep{Stevenson85}. For Saturn in particular, the H/He phase separation and helium-rain is likely to occur and could 
result in an extended inhomogeneous zone at several Mbars, and/or in a He-layer atop the core. 
  H/He phase separation in Saturn's entire interior below about 1 Mbar ($\sim 2/3$ of its radius) is supported by
modern H/He phase diagrams based on ab initio simulations \cp{Lorenzen11,Morales13}. 
Application of those new predictions to Saturn's inhomogeneous evolution remains to be done.

A different inhomogeneous model has recently been shown to provide a possible, alternative explanation 
for Saturn's high luminosity \citep{Leconte13}. In that case, the internal structure is assumed to have an 
inhomogeneous zone where the abundance of heavy elements increases with depth, and through which heat is transported 
by layered semi-convection. While details of the dynamical behavior like layer formation or merging in such a 
semi-convective zone are poorly understood, such a scenario can in principle explain an enhanced luminosity 
(the Saturn case), or a reduced luminosity (the Uranus case). In fact, by adjusting the zone's extent and the a priori 
unknown height of the convective layers, Saturn's observed luminosity can be reproduced, without requiring --albeit not excluding-- an additional energy source like helium-rain \citep{Leconte13}. A semi-convective zone may lead to a significantly higher deep internal temperatures, so that the He rain region may terminate before the core is reached. 
In the region between the semi-convective zone and the core, convection could be maintained, and the magnetic field generated.  Higher internal temperatures in the deep interior lead to a lower density for the H/He mixture, which also necessitates a larger mass of heavy elements in the planet's interior \cp{Leconte12}.  These authors find Saturn and Jupiter models with total masses of heavy elements \emph{nearly double those of adiabatic models}, including up to $\sim$30 \me\ in the H/He envelope.  Therefore, isentropic models should be thought of as providing a lower limit on heavy element content of Saturn.

New work on the influence of a double-diffusive region created by He phase separation, on the temperature gradient and cooling history of giant planets, was recently published by \ct{Nettelmann15}.  These authors couple the interior composition gradient and temperature gradient, via an iterative procedure, since there is feedback between the two gradients.  They allow the He gradient region to evolve with time, given a H/He phase diagram and a prescription for energy transport in the gradient region \cp{Mirouh12,Wood13}.  In application to Jupiter, they find that He rain can either prolong, or even \emph{shorten}, the cooling time for Jupiter to its measured \teff, depending on the efficiency of energy transport through the He-gradient region.

\begin{figure}[h!]
\includegraphics[width=0.67\textwidth]{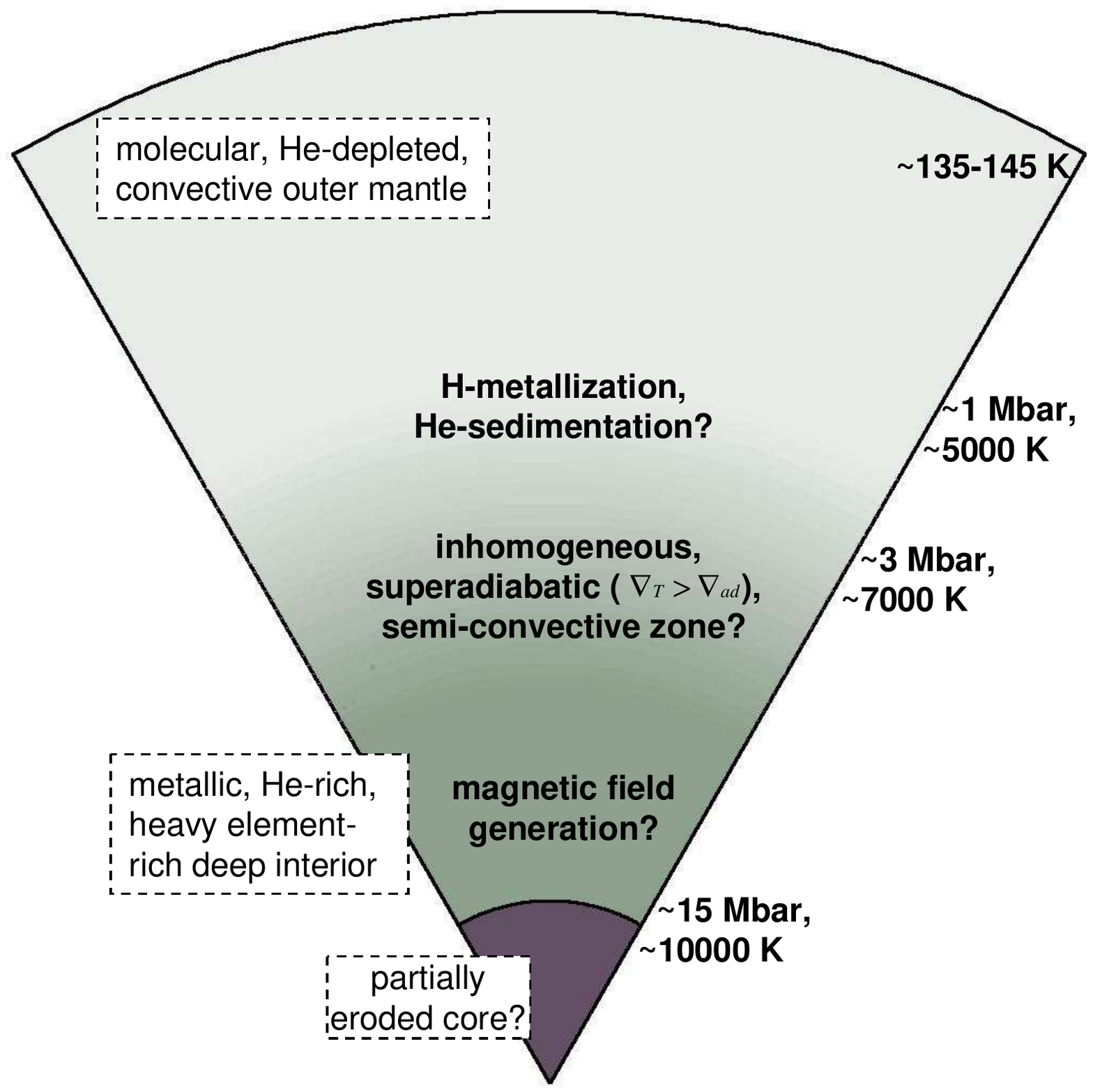}
\caption{\label{fig:nn_sat_interior}
Illustration of Saturn's possible inhomogeneous internal structure. The outer 1/3 of the planet is shown to be He-depleted, convective and homogeneous. A semi-convective region with compositional gradient (He, possibly heavy elements) separates the outer envelope from the convective, homogeneous, and metallic deep interior in which the magnetic field may be generated. Some of the initial core material may today be mixed into the deep envelope.
}
\end{figure}

Figure \ref{fig:nn_sat_interior} illustrates a possible inhomogeneous model for Saturn that could be consistent with 
its high luminosity, helium depleted atmosphere, the dipolar magnetic field, and some of the observed waves
in the rings (See section \ref{sec:seis}). 
The abundance of helium (or heavy elements) is shown to increase between 1 Mbar and 5 Mbar, as a result
of H/He phase separation and He rain in a semi-convective, superadiabatic zone. In that thick zone, non-dipolar 
moments of the magnetic field may be filtered out (see the chapter on Saturn's magnetic field), leading to a dipolar field on-top of it. Furthermore, 
the non-zero Brunt-V{\"a}is{\"a}l{\"a} frequency could allow for the generation of gravity waves that then through
mode-mixing with f-modes may explain the observed fine-splitting between the $m=-2$ modes \citep{Fuller14};  see also Section \ref{sec:seis}.

These recent developments suggest the consideration of interior models for gaseous giant planets beyond the standard assumption
of largely isentropic, homogeneous envelopes atop a well-defined core, in order to be able explain the observed properties.


\subsection{Seismology} \label{sec:seis} 

Perhaps the most straightforward method for constraining the interior properties of a planet is to study waves that propagate directly through its interior. This is the realm of seismology and on Earth both waves launched by discrete events (earthquakes) and resonant normal mode oscillations are studied. 
The detection and measurement of the frequencies of individual resonant modes trapped in the solar interior revolutionized the study of the Sun \citep{CD02} and later stars \citep{Chaplin13} and led to suggestions that the study of similar trapped
modes would open a new window into the internal structure of the giant planets \citep{Vorontsov89}. 

Several types of oscillations can be found in a fluid sphere and they are 
denoted by their primary restoring force. The most commonly discussed include 
pressure (or p-) modes that are essentially trapped
acoustic waves and gravity (or g-) modes that are resonant waves found in regions of varying static
stability. In a purely isentropic sphere, g-modes would not be present. The f-modes are the
equivalent of surface waves in a lake and are modes with no radial nodes in displacement. An 
individual 
mode is denoted by the three integers, ($\ell,\, m,\, n$) that count the total number of nodal lines on the surface, in azimuth, and in radius. Modes with $\ell = m$ are sectoral, like
segments of an orange, and the f-modes have $n=0$ by definition.

By comparing observed frequencies of these with those expected from theory the density profile throughout the interior can be tightly constrained (for a discussion of inversion methods, 
see e.g. \citet{Vorontsov89,Jackiewicz12}). Since lower order modes probe more deeply into the planet,
such modes are of the greatest interest. \ct{Gudkova95} showed that the observation of oscillation modes up to degree $\ell = 25$ would constrain both the core size and the nature
of the metallic hydrogen phase transition.
Several attempts to observe these oscillations on Jupiter were made and \citet{Schmider91} and \ct{Mosser00} reported excess oscillatory power in the expected frequency range 
for Jovian p-modes, although they were not able to identify specific modes or frequencies.  More recently, \ct{Gaulme11} detected Jupiter's oscillations, a promising first result, although the detection was not able to strongly constrain Jupiter interior models.

Telescopic searches for modes on Saturn are complicated by the rings and there have been 
no systematic surveys. However following a suggestion by \citet{Stevenson82}, 
\citet{Marley90} and \ct{Marley93} explored whether Saturn's rings might serve as a
seismograph, recording slight perturbations to the gravity field produced 
by periodic density perturbations
inside the planet. They found that even 1-m amplitude f-mode oscillations could indeed induce
perturbations to the external gravity field that in the nearby C-ring were comparable to 
those induced by distant external satellites.
The f-modes are favorable for detection because they have no radial nodes as
they perturb the density within the planet. Consequently the integrated
density perturbation from surface to deep interior is always in phase and the effect on 
the external gravitation field is greater than for any p- or g-mode, which always
have at least one radial node. Furthermore the frequencies of low
order (small $\ell$) modes serendipitously produce first order resonances in the C-ring, which lies near the planet.

Those f-modes which propagate in the same direction as Saturn rotates appear to a fixed
external observer to be of even higher frequency as their gravitational perturbation is
swept around the planet by rotation. This slight, regular, perturbation tugs on those 
ring particles 
that happen to orbit at an orbital radius where the apparent frequency of the mode is
resonant with their orbit. If the perturbation is sufficiently
large the collective response of the ring particles produces a wave feature or
even a gap in the rings. By measuring the precise location of such ring features
it would in principle be possible to infer the resonant frequency (from the orbital
dynamics) and thereby the mode frequency

After \citet{Marley91} computed new Saturn mode oscillation frequencies \citet{Marley93} 
proposed that certain wave features in Saturn's C-ring discovered by \citet{Rosen91}, as well as the Maxwell 
gap, were created by resonant interaction with low order internal f-mode oscillations of Saturn.
While \citet{Marley93} argued that the Rosen waves could be associated with saturnian oscillation modes, 
their detailed predictions for the characteristics of the waves expected to be excited in the rings could not
be tested by the \emph{Voyager} data available at the time. Almost 25 years later, however, \emph{Cassini} stellar occultation
data, obtained as stars pass behind the rings, finally allowed a test of the ring 
seismology hypothesis.

After an exhaustive analysis of the C-ring occultation data, \citet{Hedman13,Hedman14} confirmed that indeed
at least 8 unexplained C-ring wave features have the appropriate characteristics to be excited by f-mode
oscillations of Saturn. Since the precise orbital frequency is known from the location of the
wave feature, this essentially provides a very precise measurement of several specific Saturn 
oscillation frequencies, fulfilling the promise of ring seismology. However instead
of a single f-mode (with a specific $(\ell,\, m)$) being associated with the expected single C-ring feature, 
\citet{Hedman13,Hedman14}
found that two f-modes were associated with three features each. This `fine-splitting' of mode
frequencies was unexpected and is not the result of rotation alone, as the usual rotational splitting
of modes has already been accounted for in the seismological predictions. The confusion
over the multiple mode frequencies has rendered the modes value for constraining 
Saturn's interior structure somewhat problematic, at least until an appropriate
theory to explain the splitting is developed.

\citet{Fuller14} has attempted to develop such a theory. He investigated mode-mixing, where 
distinct oscillation modes can interact with one another if they have similar frequencies. He
found that the $\ell=2$ f-mode, for example, can interact with a gravity mode of Saturn if there is a convectively stable region above Saturn's core. In essence the f-mode and gravity mode interact and the result is
a mode of mixed character that splits the $\ell = 2$ f-mode oscillation frequency. If this is indeed 
the cause of the observed splitting then this may be offering a precise measure of
core erosion (or a deep He-gradient) in the deep interior of Saturn. Combined with the other measured mode frequencies,
seismology may hold promise for constraining not only the size of Saturn's core, but also the deep
composition of the planet. More theoretical development is required, however, to fully exploit this
opportunity. Nevertheless ring seismology likely now has great promise for opening a unique window 
into Saturn's interior structure.

\subsection{Future Prospects}
\subsubsection{Cassini Grand Finale} 
\begin{figure*}[h!]
\centering
\includegraphics[width=0.8\textwidth]{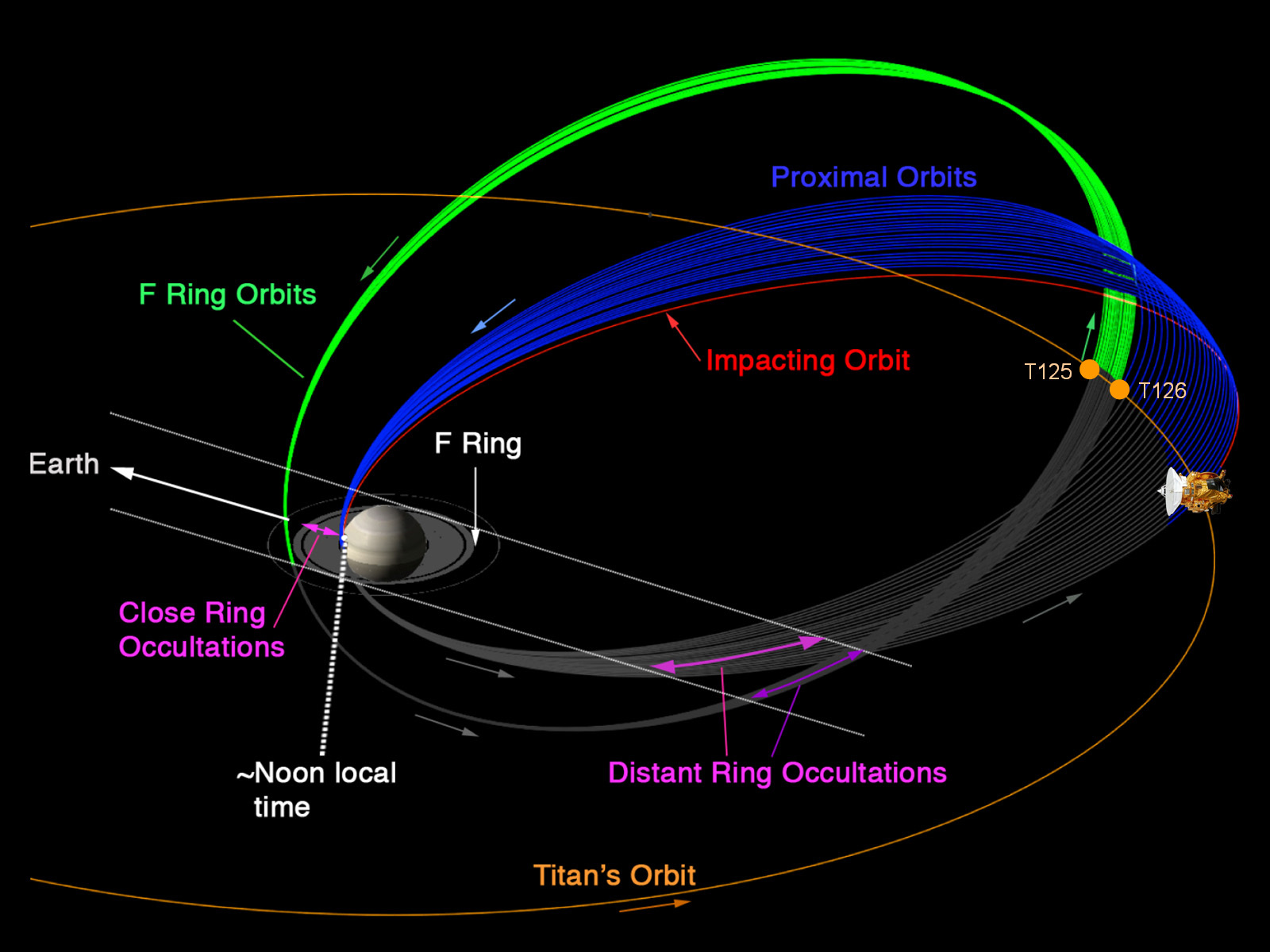}
\caption{\label{fig:orbits}
The \emph{Cassini} Grand Finale orbits. After completion of the first 20 F ring orbits, \emph{Cassini} will pass 22 times between the inner edge of the D ring and Saturn's atmosphere, with an orbital inclination of 63.4 degrees and a pericenter latitude between 5.5 and 7.5 degrees south. The final plunge into Saturn's atmosphere, required by planetary protection rules, is currently scheduled for Sept. 15, 2017. 
}
\end{figure*}

Before the planned termination and plunge into Saturn's atmosphere in September 2017, the \emph{Cassini} mission will execute 20 orbits with a 7.2 day period and pericenter at about 2.5 Saturn radii, and 22 highly inclined (63.4 degree) orbits with a period of 6.2 day and a periapsis altitude about 1700 km above the 1-bar pressure level. (see Figure \ref{fig:orbits}). This set of orbits, named the \emph{Cassini} Grand Finale, have been tailored to carry out close observations of Saturn and to probe its interior structure by means of gravity and magnetic field measurements. Although neither the spacecraft nor its instruments were designed for this kind of observations, the scientific return is expected to be high. Thanks to the proximity of the spacecraft to Saturn in the final 22 orbits (just inside the inner edge of the D ring), \emph{Cassini} will return the harmonic coefficients of the magnetic and gravity field to about degree 10 or larger. 

While the onboard magnetometer will carry out continuous measurements throughout the last 22 orbits, starting from April 2017, there are much fewer opportunities for gravity measurements, which are only possible when the high gain antenna is Earth-pointed. Operational constraints (such as the elevation angle at the ground station and the protection of the spacecraft from dust hazard using the antenna as a shield), and the need to share the observation time between the onboard instruments, drastically reduce to six the number of orbits devoted to gravity science. 

The determination of Saturn's gravity field will be carried out by means of range rate measurements and sophisticated orbit determination codes. Range rate is routinely measured at a ground antenna by transmitting a highly stable monochromatic microwave signal to the spacecraft. An onboard transponder receives the signal and coherently retransmits it back to ground, where the Doppler shift between the outgoing and incoming signal is measured. The antennas of NASA's Deep Space Network (DSN), operating at X-band (7.2-8.4 GHz), provide range rate accuracies of about 12 micron/s at 1000 s integration times, and about a factor of four larger at a time scale of 60 s. Thanks to the use of higher frequencies (32-34 GHz), and the consequent immunity to propagation noise from interplanetary plasma, a similar experiment on the \emph{Juno} mission will exploit observable quantities about a factor of four less noisy. 

In spite of the limited number of orbits, \emph{Cassini}'s gravity measurements will take advantage of a remarkably favorable orbital geometry, which always ensures a large projection of the spacecraft velocity along the line of sight when the spacecraft is close to the planet. Numerical simulations of the gravity experiment in a realistic scenario indicate that \cas\ will be able to estimate the (unnormalized) zonal harmonics with accuracies ranging from $2\times10^{-9}$ for $J_2$ and $1.5\times10^{-7}$ for $J_{12}$.  This precision will provide unprecedented constraints on the density structure of the outer H/He envelope, and allow for the detection of the depth of zonal flows seen in the visible atmosphere \cp[e.g.][]{Kaspi10,Kaspi13}.  
It is expected that the \cas\ data will provide also an estimate of Saturn's $k_2$ and $k_3$ Love numbers to accuracies respectively of 0.015 and 0.12, and the gravitational parameter $GM$ of the B ring to 0.15 km$^{3}$ s$^{-2}$.  

\subsubsection{What is needed for future progress}
What has emerged from the still unfolding \cas\ era is a picture of Saturn's interior that is full of complexity.  The completion of the \cas\ revolution will utilize the tremendously more precise data on the planet's gravity field and magnetic field that are essentially assured from the \emph{Cassini} Grand Finale orbits.  To maximize the science from these unique data sets will require a concomitant signifant push in the analysis of existing \cas\ data sets, laboratory studies, and theoretical work.  Below we suggest areas for future work.
\begin{packed_enum}
   \item Better knowledge of the phase diagram for helium immiscibility.  The most recent \emph{ab initio} phase diagrams \cp{Lorenzen11,Morales13} yield similar, but clearly discrepant predictions for the temperatures of the onsent of phase separation.  While it appears that most of Saturn's interior is in a region with a gradient in helium abundance, additional theoretical and experimental work are needed to bring confidence to our understanding of the phase diagram.
   \item A determination of Saturn's atmospheric He abundance from \emph{Cassini} spectroscopic and occultation observations.  This is a complicated issue \cp[e.g.][]{CG00}, but uncertainty in this value will long dominate our uncertain knowledge of Saturn's cooling history.  In tandem, a derivation of atmospheric $P-T$ profile, including the 1-bar temperature, would put our knowledge of the planet's thermal profile on more solid footing.
   \item Three dimensional simulations of the transport of helium and energy within the helium immiscibility region.  Recent work \cp{Wood13} on double-diffusive convection is an important step in this direction, but the coupled nature of the helium and temperature gradients within giant planets warrants additional focused work.
   \item If the first three items are addressed, a new generation of Saturn thermal evolution models is certainly warranted, which could simultaneously match the planet's intrinsic flux, 1-bar temperature, and current atmospheric helium abundance.
   \item The most dramatic advance in our understanding of the planets's interior would surely come from a wider exploration of the range of interior structures that are consistent with both the gravity field and seismology data.  \cite{Fuller14} have begun this work with a limited range of Saturn interior models.  However, a wider exploration of the utility of the seismology data sets, in tandem with the Grand Finale gravity field constraints, is clearly needed.
\item Measurements from \emph{Cassini}'s Solstice mission and contemporary model improvements will not close our current knowledge gaps on fundamental properties of Saturn and our knowledge of giant planet formation in the solar system. In line with \cite{Mousis2014} we suggest future space exploration of Saturn by means of an entry probe into its atmosphere in order to determine accurately the abundances and isotopic ratios of noble gases. 
Such values contain unique information not only on internal properties such as helium rain and bulk composition but also on the early history of the solar system.  

\end{packed_enum}

\small{

}

\newpage

\begin{figure*}[h]
\centering
\includegraphics[width = 6.5 in]{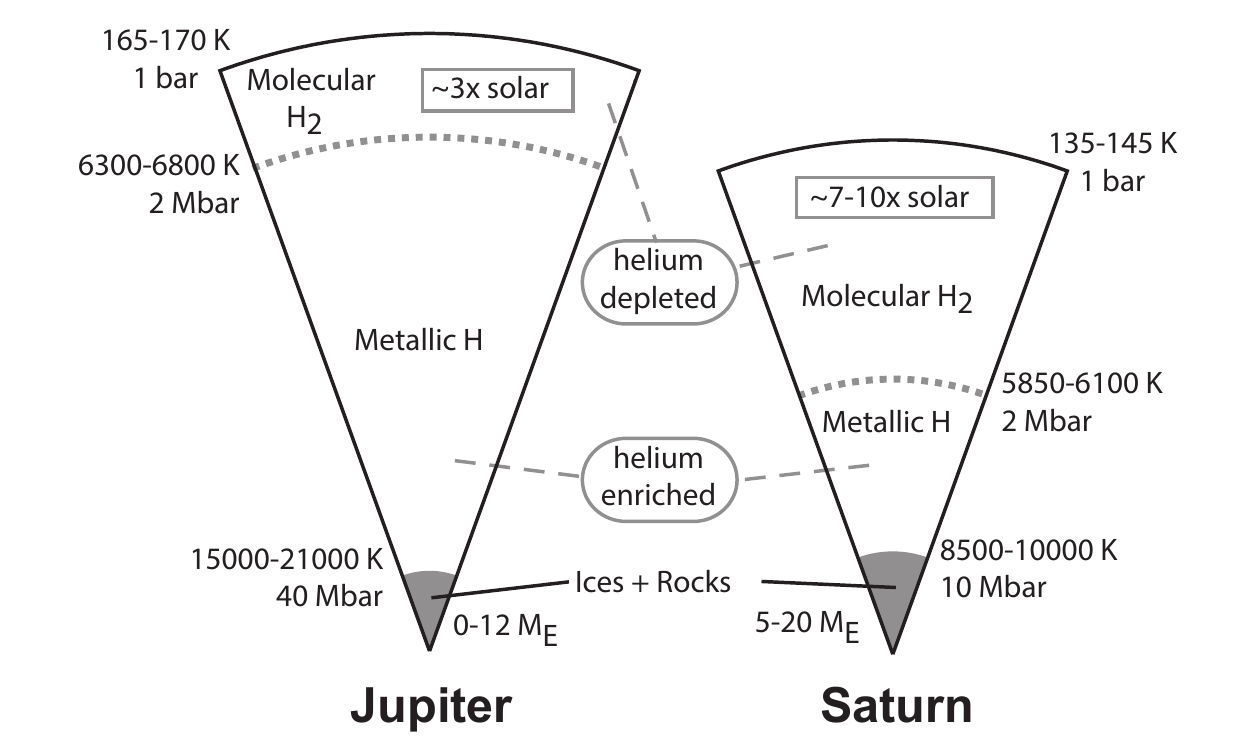}
\caption{Highly idealized comparative view on the interiors of Jupiter and Saturn.}
\label{fig:compare}
\end{figure*}

\end{document}